\documentclass[manuscript]{aastex}

\newcommand{\cmda}{Celestial Mechanics and Dynamical Astronomy}
\newcommand{\pepi}{Physics of the Earth and Planetary Interiors}

\shorttitle{Librations of a synchronous satellite}
\shortauthors{Noyelles}

\usepackage{lscape}
\usepackage{color}
\usepackage{hyperref}
\usepackage{wasysym}
\begin{document}


\title{Interpreting the librations of a synchronous satellite -- How their phase assesses Mimas' global ocean}


\author{Beno\^it Noyelles}
\affil{NAmur Center for CompleX SYStems (NAXYS) -- University of Namur -- Rempart de la Vierge 8 -- B-5000 Namur -- Belgium}
\email{benoit.noyelles@unamur.be}



\begin{abstract}

\par Most of the main planetary satellites of our Solar System are expected to be in synchronous rotation, the departures from the strict 
synchronicity being a signature of the interior. Librations have been measured for the Moon, Phobos, and some satellites of Saturn. I here
revisit the theory of the longitudinal librations in considering that part of the interior is not hydrostatic, i.e. has not been shaped
by the rotational and tidal deformations, but is fossil. This consideration affects the rotational behavior. 

\par For that, I derive the tensor of inertia of the satellite in splitting these two parts, before proposing an analytical solution
that I validate with numerical simulations. I apply this new theory on Mimas and Epimetheus, for which librations have been measured
from Cassini data. 

\par I show that the large measured libration amplitude of these bodies can be explained by an excess of triaxiality that would not
result from the hydrostatic theory. This theory cannot explain the phase shift which has been measured in the diurnal librations
of Mimas. This speaks against a solid structure for Mimas, i.e. Mimas could have a global internal ocean.

\end{abstract}





\section{Introduction}

\par The Cassini space mission has given us invaluable data on the satellites of Saturn. In particular, we now dispose of 
measurements of their shapes as triaxial ellipsoids \citep[e.g.]{t2010}, and the rotation of Janus, Epimetheus \citep{ttb2009}, Mimas \citep{trlcrrn2014}, 
Enceladus \citep{tttbjlhp2016}, and Titan \citep{sklhloacgiphjw2008,mislm2016} have been measured. An issue is to get clues on the interior from these informations.

\par Most of the natural satellites of the giant planets are expected to have reached a state of synchronous rotation, known as Cassini State. This is a dynamical equilibrium
from which small departures are signatures of the internal structure. These small departures are longitudinal and latitudinal librations, the latter ones translating into an obliquity.
The main part of the longitudinal librations is a periodic diurnal oscillation, named physical librations. For a rigid body, they read \citep[Eq. 5.123, e.g.]{md2000}

\begin{equation}
\label{eq:physrigid}
\gamma(t) = \frac{2e}{1-\left(n/\omega_0\right)^2}\sin(nt),
\end{equation}
$e$ being the orbital eccentricity of the satellite, $n$ its orbital frequency, and $\omega_0$ the frequency of the small proper oscillations around the equilibrium. 
We have:

\begin{equation}
\label{eq:omegurig}
\omega_0 = n\sqrt{3\frac{I_{22}-I_{11}}{I_{33}}G_{200}(e)},
\end{equation}
where the quantities $I_{xx}$ are diagonal elements of the tensor of inertia of the satellite, and $G_{200}(e)$
is an eccentricity function made popular by \citet{k1966}:

\begin{equation}
\label{eq:kaulag200}
G_{200}(e) = 1-\frac{5}{2}e^2+\frac{13}{16}e^4-\frac{35}{288}e^6+\mathcal{O}(e^8).
\end{equation}
This function is also known as the Hansen function $H(1,e)$, and is present in \citet{c1861}.

\par The quantity $(I_{22}-I_{11})/I_{33}$, sometimes written as $(B-A)/C$, represents the equatorial ellipticity, or the triaxiality, of
the distribution of mass inside the satellite. This formula assumes a rigid shape, i.e. the tensor of inertia is constant.

\par A seminal paper by \citet{gm2010} has shown that if the satellite is not strictly rigid, but viscoelastic,
then a restoring torque tends to counterbalance the tidal torque of the parent planet, to lower the amplitude
of the physical librations. This has motivated recent studies \citep{vbt2013,rrc2014,mfd2016}, which revisit the theory 
of the librations in including a tidal parameter, $k_2$ or $h_2$, which characterizes the amplitude of variation
of the gravity field, or of the surface, at the diurnal, or orbital, frequency $n$. In these studies, the body is assumed to be
at the hydrostatic equilibrium.

\par The theory of the hydrostatic equilibrium tells us that the mass distribution in the body, i.e. its inertia, is shaped by its rotation and the tidal torque of its parent planet,
while the inertia rules its rotation. However, in most of the studies, the inertia is mainly composed of a constant component which has no chance to be shaped by the
rotation, while being assumed to correspond to the hydrostatic equilibrium.

\par In this paper I go further, in assuming that part of the mass distribution of these bodies is frozen, while part of it is still being shaped by
the rotational and tidal deformation. For that, I first show from the measured radii that the departures from the hydrostatic equilibrium are ubiquitous 
in the system of Saturn (Sect.~\ref{sec:nonhydro}). Then I express the tensor of inertia of a satellite orbiting a giant planet, in considering
a frozen triaxiality superimposed with elastic deformation (Sect~\ref{sec:inertia}). I then deduce the librational dynamics of the satellite (Sect.~\ref{sec:librations}), which I apply to the specific cases of Epimetheus and Mimas, for which longitudinal
librations have been measured, and which are assumed to have rigid structure. Finally, I introduce the dissipative part of the tides (Sect.~\ref{sec:dissipative}), to investigate their
influence on the measurements. The reader can refer to Tab.~\ref{tab:notations} for the notations.

\section{Departures from the hydrostatic equilibrium\label{sec:nonhydro}}

\par Usually the shape of such a body is assumed to be the signature of an hydrostatic equilibrium. This means that the mass distribution
is an equilibrium between the gravity of the body, and the rotational and tidal deformations it is subjected to. For a natural satellite
orbiting a giant planet, it is assumed that the tidal deformation is only due to the planet, and that its spin rate is equal to its
orbital rate, i.e. it rotates synchronously. This gives the following relations \citep{mn2009,cr2013,t2014}:

\begin{eqnarray}
\frac{\mathfrak{b}-\mathfrak{c}}{\mathfrak{a}-\mathfrak{c}} \approx \frac{1}{4}-\frac{495}{224}q+\frac{51}{32}e^2, \label{eq:hydroshape} \\
\frac{J_2}{C_{22}} \approx \frac{10}{3}-\frac{400}{21}q+\frac{34}{3}e^2, \label{eq:hydrograv}
\end{eqnarray}
$\mathfrak{a}>\mathfrak{b}>\mathfrak{c}$ being respectively the 3 planet-facing, orbit-facing, and polar radii, $J_2$ and $C_{22}$ the classical Stokes coefficients, and 
$q=n^2R^3/(\mathcal{G}M)$. In this last formula, $M$ is the mass of the satellite, and $R$ its mean radius. The relation (\ref{eq:hydroshape}) holds for the shape, and (\ref{eq:hydrograv}) for the gravity field. The synchronous rotation results in a
forcing of the triaxiality of the satellite, since it on average always presents the same face to its parent planet.

\par The Cassini mission has given us invaluable data on the shapes (Tab.~\ref{tab:shapes}) and gravity fields (Tab.~\ref{tab:gravity}) of some Saturnian satellites,
and a comparison with the numbers predicted by the hydrostatic theory reveals some departure (Tab.~\ref{tab:nonhydro}), as for the Moon 
\citep[e.g.]{j1937,lp1980,gpnz2014} and Phobos \citep[e.g.]{lrrcdm2013}.

\placetable{tab:gravity}

\placetable{tab:shapes}

\placetable{tab:nonhydro}

\par These departures from the hydrostatic equilibrium may have different origins, such as a fossil shape, internal processes,
or impacts\ldots In the following, I consider them as frozen inertia.

\section{The tensor of inertia\label{sec:inertia}}

\par I consider a homogeneous, triaxial and synchronous satellite. Its orbital dynamics comes from the 
oblate two-body problem, i.e. the semimajor axis $a$ and the eccentricity $e$ are constant, and the mean longitude $\lambda$ and longitude of the pericentre $\varpi$ have a 
uniform precessional motion, the associated frequencies being $n$ and $\dot{\varpi}$, respectively. Moreover, its orbit is assumed to lie in the equatorial plane of the planet,
as a consequence the satellite has no obliquity. I also neglect the polar motion, the angular momentum of the satellite being collinear to its polar axis of inertia.

\par The tensor of inertia of the satellite $\mathcal{I}$ can be decomposed as:

\begin{equation}
\label{eq:wholetensor}
\mathcal{I} = \mathcal{I}^{(f)}+\mathcal{I}^{(s)}+\mathcal{I}^{(t)}+\mathcal{I}^{(r)},
\end{equation}
where 

\begin{itemize}

\item $\mathcal{I}^{(f)}$ is the frozen component. It is constant and its physical origin is here not addressed,

\item $\mathcal{I}^{(s)}$ is the inertia of a spherical body. We have

\begin{equation}
\label{eq:Isth}
I_{ij}^{(s)} = \iiint_{\mathrm{body}} \rho(r) r^2\,\textrm{d}V\delta_{ij},
\end{equation}
which gives, for our homogeneous body: 

\begin{equation}
\label{eq:Is}
I_{ij}^{(s)} = \frac{2}{5}MR^2\delta_{ij},
\end{equation}
$\delta_{ij}$ being the classical Kronecker symbol, $\rho(r)$ the density at the distance $r$, and $\textrm{d}V$ a volume element.

\item $\mathcal{I}^{(t)}$ is the deformation of the inertia induced by the tides, and

\item $\mathcal{I}^{(r)}$ is the rotational deformation.

\end{itemize}

\par In this section, I neglect the dissipative part of the deformation. 
As a consequence, the sum $\mathcal{I}^{(t)}+\mathcal{I}^{(r)}$ represents an elastic deformation, and will be denoted $\mathcal{I}^{(e)}$, e standing for \emph{elastic}. 
The dissipation will be considered in the Sect.~\ref{sec:dissipative}.

\subsection{The frozen inertia}

\par The frozen part of the tensor of inertia is constant. As a symmetric tensor, it can be written under a diagonal form in an appropriate reference 
frame $(\hat{f_1},\hat{f_2},\hat{f_3})$, whose axes are the principal axes of inertia, i.e.:

\begin{equation}
\label{eq:tensorfroz}
\mathcal{I}^{(f)} = \left(\begin{array}{ccc}
I_{11}^{(f)} & 0 & 0 \\
0 & I_{22}^{(f)} & 0 \\
0 & 0 & I_{33}^{(f)}
\end{array}\right),
\end{equation}
with $I_{11}^{(f)} \leq I_{22}^{(f)} \leq I_{33}^{(f)}$. In the literature dealing with rigid rotation, these 3 moments of inertia are often written $A$, $B$ and $C$, respectively.

\subsection{The elastic inertia}

\par The elastic tensor $\mathcal{I}^{(e)}$ results from the combined action of the tidal torque of the parent planet and the rotation of the 
satellite. Following \citep[e.g.]{wbyrd2001}, we have

\begin{equation}
\label{eq:It}
I_{ij}^{(t)} = -k_2\frac{M_{\saturn}R^5}{r^3}\left(x_ix_j-\frac{\delta_{ij}}{3}\right),
\end{equation}
and

\begin{equation}
\label{eq:Ir}
I_{ij}^{(r)} = \frac{k_2R^5}{3\mathcal{G}}\left(\omega_i\omega_j-\frac{n^2}{3}\delta_{ij}\right),
\end{equation}
where $x_1=x$, $x_2=y$ and $x_3=z$ are the coordinates of the unit vector pointing to the parent planet in the reference frame of the 
principal axes of inertia of the satellite. In this frame, $\vec{\omega} = \omega_1\hat{f_1}+\omega_2\hat{f_2}+\omega_3\hat{f_3}$ is
the rotation vector of the satellite. $k_2$ is the second order gravitational Love number, which characterizes the amplitude of deformation
of the gravity field. $M_{\saturn}$ is the mass of the parent planet, $R$ is the mean radius of the satellite, and $r$ is the distance
between the barycentres of the planet and the satellite.

\par Our assumptions imply $x_3 = 0$, $\omega_1=\omega_2=0$ and $\omega_3=n$, this results in:

\begin{eqnarray}
I_{11}^{(e)} & = & -k_2\frac{M_{\saturn}R^5}{3a^3}\left(1+\left(\frac{a}{r}\right)^3\left(2x^2-y^2\right)\right), \label{eq:I11cart}  \\
I_{12}^{(e)} & = & -k_2\frac{M_{\saturn}R^5}{r^3}xy, \label{eq:I12cart} \\
I_{22}^{(e)} & = & -k_2\frac{M_{\saturn}R^5}{3a^3}\left(1+\left(\frac{a}{r}\right)^3\left(2y^2-x^2\right)\right), \label{eq:I22cart}  \\
I_{33}^{(e)} & = & k_2\frac{M_{\saturn}R^5}{9a^3}\left(2+3\left(\frac{a}{r}\right)^3\right), \label{eq:I33cart} \\
I_{13}^{(e)} & = & I_{23}^{(e)} = 0, \label{eq:I123zero}
\end{eqnarray}
$a$ being the semimajor axis of the satellite.

\par The coordinates of the parent planet $x$ and $y$ can now be expressed with respect to the orbital elements of the satellite and to its orientation.
Since it is assumed to have only a longitudinal motion, its orientation is given by an instantaneous rotation angle $p$ around $\hat{f_3}$ such that
$\dot{p}=n$ and its origin $p=0$ corresponds to a state in which the long axis of the satellite points to the barycenter of the parent planet.

In the reference frame $(\hat{f_1},\hat{f_2},\hat{f_3})$, we have

\begin{equation}
\label{eq:xyz}
\left(\begin{array}{c}
x \\
y \\
z \end{array}\right) = \left(\begin{array}{ccc}
\cos p & \sin p & 0 \\
-\sin p & \cos p & 0 \\
0 & 0 & 1 \end{array}\right)\times \left(\begin{array}{c}
\cos (f+\varpi) \\
\sin (f+\varpi) \\
0 \end{array}\right),
\end{equation}
where $f$ stands for the true anomaly, and $\varpi$ for the longitude of the pericenter. These quantities can be introduced in the formulae (\ref{eq:I11cart}) to (\ref{eq:I33cart}), before being expanded with 
respect to the eccentricity $e$ using the classical formulae \citep[Eq.~3.117]{d2002}, \citep[Eq.~2.84 \& 2.85]{md2000}:

\begin{eqnarray}
\frac{a}{r} & = & 1+2\sum_{\nu=1}^{\infty} J_{\nu}(\nu e)\cos (\nu(\lambda-\varpi)), \label{eq:asr} \\
\cos f & = & 2\frac{1-e^2}{e}\sum_{\nu=1}^{\infty} J_{\nu}(\nu e)\cos (\nu(\lambda-\varpi))-e, \label{eq:cosf} \\
\sin f & = & 2\sqrt{1-e^2}\sum_{\nu=1}^{\infty}\frac{\textrm{d}J_{\nu}(\nu e)}{\textrm{d}e}\frac{\sin(\nu(\lambda-\varpi))}{\nu}, \label{eq:sinf}
\end{eqnarray}
where $J_{\nu}$ are the Bessel functions of the first kind. I now introduce the argument of the synchronous spin-orbit resonance $\sigma=p-\lambda+\pi$ \citep[e.g.]{h2005}
and the mean anomaly $\mathcal{M}=\lambda-\varpi$ to get, up to the second order in the eccentricity:

\begin{eqnarray}
I_{11}^{(e)} & = & k_2\frac{M_{\saturn}R^5}{a^3}\left(-\frac{5}{18}-\frac{e^2}{4}-\frac{1}{2}\left(1-\frac{5}{2}e^2\right)\cos 2\sigma\right. \nonumber \\
& & \left.+\frac{e}{4}\left(\cos(\mathcal{M}+2\sigma)-2\cos(\mathcal{M})-7\cos(\mathcal{M}-2\sigma)\right)\right. \nonumber \\
& & \left.-\frac{e^2}{4}\left(3\cos(2\mathcal{M})+17\cos(2\mathcal{M}-2\sigma)\right)\right), \label{eq:I11orbit} \\
I_{12}^{(e)} & = & \frac{k_2}{2}\frac{M_{\saturn}R^5}{a^3}\left(\left(1-\frac{5}{2}e^2\right)\sin 2\sigma-\frac{e}{2}\left(\sin(\mathcal{M}+2\sigma)+7\sin(\mathcal{M}-2\sigma)\right)\right. \nonumber \\
& & \left.-\frac{17}{2}e^2\sin(2\mathcal{M}-2\sigma)\right), \label{eq:I12orbit} \\
I_{22}^{(e)} & = & k_2\frac{M_{\saturn}R^5}{a^3}\left(-\frac{5}{18}-\frac{e^2}{4}+\frac{1}{2}\left(1-\frac{5}{2}e^2\right)\cos 2\sigma\right. \nonumber \\
& & \left.+\frac{e}{4}\left(7\cos(\mathcal{M}-2\sigma)-\cos(\mathcal{M}+2\sigma)-2\cos(\mathcal{M})\right)\right. \nonumber \\
& & \left.+\frac{e^2}{4}\left(17\cos(2\mathcal{M}-2\sigma)-3\cos(2\mathcal{M})\right)\right), \label{eq:I22orbit} \\
I_{33}^{(e)} & = & k_2\frac{M_{\saturn}R^5}{a^3}\left(\frac{5}{9}+\frac{1}{2}e^2+e\cos(\mathcal{M})+\frac{3}{2}e^2\cos(2\mathcal{M})\right). \label{eq:I33orbit} 
\end{eqnarray}

\par The argument of the spin-orbit resonance $\sigma$ is usually set to $0$, except in \citep[Eq.~37-38]{cr2013}. Here, I keep $\sigma$ as a variable, since it will be involved in the
equation of the librations. Since the rotation is assumed to shape the inertia, and the inertia rules the rotation, then $\sigma$ should be let free to vary.
I will anyway assume that the frequency associated with $\sigma$ is null, i.e. I assume $\sigma$ to be constant on average, which is consistent with the resonant locking. 
The formulae (\ref{eq:I11orbit}) to (\ref{eq:I33orbit}) imply that 
the tidal response of the satellite does not depend on the frequency of the excitation. Actually the tidal parameter is frequency-dependent. To model this dependency I define the following
frequencies:

\begin{eqnarray}
\nu_0 & = & 0, \label{eq:nu0} \\
\nu_1 & = & |n-\dot{\varpi}|, \label{eq:nu1} \\
\nu_2 & = & |2n-2\dot{\varpi}|, \label{eq:nu2}
\end{eqnarray}
and the elastic tensor of inertia becomes:

\begin{eqnarray}
I_{11}^{(e)} & = & \frac{M_{\saturn}R^5}{a^3}\left(k_2(\nu_0)\left(-\frac{5}{18}-\frac{e^2}{4}-\frac{1}{2}\left(1-\frac{5}{2}e^2\right)\cos 2\sigma\right)\right. \nonumber \\
& & \left.+k_2(\nu_1)\frac{e}{4}\left(\cos(\mathcal{M}+2\sigma)-2\cos(\mathcal{M})-7\cos(\mathcal{M}-2\sigma)\right)\right. \nonumber \\
& & \left.-k_2(\nu_2)\frac{e^2}{4}\left(3\cos(2\mathcal{M})+17\cos(2\mathcal{M}-2\sigma)\right)\right), \label{eq:I11ortid} \\
I_{12}^{(e)} & = & \frac{M_{\saturn}R^5}{2a^3}\left(k_2(\nu_0)\left(1-\frac{5}{2}e^2\right)\sin 2\sigma-k_2(\nu_1)\frac{e}{2}\left(\sin(\mathcal{M}+2\sigma)+7\sin(\mathcal{M}-2\sigma)\right)\right. \nonumber \\
& & \left.-\frac{17}{2}e^2k_2(\nu_2)\sin(2\mathcal{M}-2\sigma)\right), \label{eq:I12ortid} \\
I_{22}^{(e)} & = & \frac{M_{\saturn}R^5}{a^3}\left(k_2(\nu_0)\left(-\frac{5}{18}-\frac{e^2}{4}+\frac{1}{2}\left(1-\frac{5}{2}e^2\right)\cos 2\sigma\right)\right. \nonumber \\
& & \left.+\frac{e}{4}k_2(\nu_1)\left(7\cos(\mathcal{M}-2\sigma)-\cos(\mathcal{M}+2\sigma)-2\cos(\mathcal{M})\right)\right. \nonumber \\
& & \left.+\frac{e^2}{4}k_2(\nu_2)\left(17\cos(2\mathcal{M}-2\sigma)-3\cos(2\mathcal{M})\right)\right), \label{eq:I22ortid} \\
I_{33}^{(e)} & = & \frac{M_{\saturn}R^5}{a^3}\left(k_2(\nu_0)\left(\frac{5}{9}+\frac{1}{2}e^2\right)+ek_2(\nu_1)\cos(\mathcal{M})+\frac{3}{2}e^2k_2(\nu_2)\cos(2\mathcal{M})\right). \label{eq:I33ortid} 
\end{eqnarray}

In the literature, $k_2(\nu_0)$ is sometimes denoted as the fluid, or secular, Love number $k_f$. It represents an indefinitely slow deformation. However,
$k_2(\nu_1)$ and $k_2(\nu_2)$ are often assumed to be equal and denoted as $k_2$. The periods of the deformations associated are respectively the orbital and
half the orbital ones, which are also the diurnal and semi-diurnal periods of the satellite.

\section{The librational dynamics\label{sec:librations}}

\subsection{The librational equations}

\par The gravitational torque of the parent planet, which acts on the satellite, is collinear to the polar axis, since we consider a planar orbit. Its non-null component 
$\Gamma$ reads \citep[e.g.]{wbyrd2001}:

\begin{equation}
\label{eq:Gammaz2}
\Gamma = 3\frac{\mathcal{G}M_{\saturn}}{r^3}\left((I_{22}-I_{11})xy+I_{12}(x^2-y^2)\right).
\end{equation}

\par After expansion with respect to the orbital elements, I get, up to the second order in the eccentricity:

\begin{eqnarray}
\Gamma & = & \left(-\frac{3}{2}+\frac{15}{4}e^2\right)\left(I_{22}-I_{11}\right)^{(f)}n^2\sin2\sigma+\frac{3}{4}\left(I_{22}-I_{11}\right)^{(f)}en^2\left(\sin(\mathcal{M}+2\sigma)+7\sin(\mathcal{M}-2\sigma)\right) \nonumber \\
 & & +\frac{51}{4}\left(I_{22}-I_{11}\right)^{(f)}e^2n^2\sin(2\mathcal{M}-2\sigma)+6en^2\frac{M_{\saturn}R^5}{a^3}\left(k_f-k_2(\nu_1)\right)\sin\mathcal{M} \nonumber \\
& & +\frac{51}{4}n^2e^2\frac{M_{\saturn}R^5}{a^3}\left(k_f-k_2(\nu_2)\right)\sin2\mathcal{M}, \label{eq:Gammadev}
\end{eqnarray}
while $\Gamma$ is also the time-derivative of the norm of the angular momentum. This yields

\begin{equation}
\label{eq:thcinetiq}
\Gamma = \frac{d(I_{33}\dot{p})}{dt},
\end{equation}
i.e.

\begin{eqnarray}
\Gamma & = & \left(\frac{2}{5}MR^2+\frac{M_{\saturn}R^5}{a^3} \left(k_f\left(\frac{5}{9}+\frac{1}{2}e^2\right)+ek_2(\nu_1)\cos\mathcal{M}+\frac{3}{2}e^2k_2(\nu_2)\cos2\mathcal{M}\right)\right)\ddot{\sigma} \nonumber \\
 & & -\frac{M_{\saturn}R^5}{a^3}(n-\dot{\varpi})\left(k_2(\nu_1)e\sin\mathcal{M}+3k_2(\nu_2)e^2\sin2\mathcal{M}\right)\dot{\sigma}.\label{eq:thcinetiq2}
\end{eqnarray}
\par The frozen component of $I_{33}$, i.e. $I_{33}^{(f)}$, should appear in this equation, but since $I_{33}$ is not affected by $\sigma$, then its secular part has the same behavior
than its frozen part, distinguishing it is in that specific case useless. So, it is kind of absorbed in $\frac{2}{5}MR^2+\frac{M_{\saturn}R^5}{a^3} k_f\left(\frac{5}{9}+\frac{1}{2}e^2\right)$.

\par In equating the formula (\ref{eq:Gammadev}) with the (\ref{eq:thcinetiq2}), I get the differential equation ruling the longitudinal libration of the satellite, i.e.

\begin{eqnarray}
& & K_1\sin2\sigma+K_2\sin(\mathcal{M}+2\sigma)+K_3\sin(\mathcal{M}-2\sigma)+K_4\sin(2\mathcal{M}-2\sigma)+K_5\sin\mathcal{M}+K_6\sin2\mathcal{M} \nonumber \\
& = & K_7\ddot{\sigma}+K_8\ddot{\sigma}\cos\mathcal{M}+K_9\ddot{\sigma}\cos2\mathcal{M}+K_{10}\dot{\sigma}\sin\mathcal{M}+K_{11}\dot{\sigma}\sin2\mathcal{M} \label{eq:biglibra}
\end{eqnarray}
with

\begin{eqnarray}
K_1    & = & \left(-\frac{3}{2}+\frac{15}{4}e^2\right)(I_{22}-I_{11})^{(f)}n^2, \label{eq:K1} \\
K_2    & = & \frac{3}{4}(I_{22}-I_{11})^{(f)}en^2, \label{eq:K2} \\
K_3    & = & \frac{21}{4}(I_{22}-I_{11})^{(f)}en^2=7K_2, \label{eq:K3} \\
K_4    & = & \frac{51}{4}(I_{22}-I_{11})^{(f)}e^2n^2, \label{eq:K4} \\
K_5    & = & 6en^2\frac{M_{\saturn}R^5}{a^3}\left(k_f-k_2(\nu_1)\right), \label{eq:K5} \\
K_6    & = & \frac{51}{4}e^2n^2\frac{M_{\saturn}R^5}{a^3}\left(k_f-k_2(\nu_2)\right), \label{eq:K6} \\
K_7    & = & \frac{2}{5}MR^2+\frac{M_{\saturn}R^5}{a^3}k_f\left(\frac{5}{9}+\frac{e^2}{2}\right), \label{eq:K7} \\
K_8    & = & \frac{M_{\saturn}R^5}{a^3}ek_2(\nu_1), \label{eq:K8} \\
K_9    & = & \frac{3}{2}\frac{M_{\saturn}R^5}{a^3}e^2k_2(\nu_2), \label{eq:K9} \\
K_{10} & = & -\frac{M_{\saturn}R^5}{a^3}(n-\dot{\varpi})ek_2(\nu_1), \label{eq:K10} \\
K_{11} & = & -3\frac{M_{\saturn}R^5}{a^3}(n-\dot{\varpi})e^2k_2(\nu_2). \label{eq:K11}
\end{eqnarray}

\subsection{The librational solution}

\par The equation (\ref{eq:biglibra}) can be simplified for an analytical resolution in neglecting the variations of $\sigma$ 
with respect to the ones of $\mathcal{M}$, and $\dot{\sigma}$. It becomes:

\begin{equation}
\label{eq:smalllibra}
\ddot{\sigma}+\omega_0^2\sigma=\kappa_1\sin\mathcal{M},
\end{equation}
with 

\begin{eqnarray}
\omega_0^2 & = & \left(3-\frac{15}{2}e^2\right)n^2\frac{(I_{22}-I_{11})^{(f)}}{\frac{2}{5}MR^2+k_f\left(\frac{5}{9}+\frac{e^2}{2}\right)M_{\saturn}\frac{R^5}{a^3}}, \label{eq:omega02} \\
\kappa_1 & = & 6en^2\frac{(I_{22}-I_{11})^{(f)}+M_{\saturn}\frac{R^5}{a^3}\left(k_f-k_2(\nu_1)\right)}{\frac{2}{5}MR^2+k_f\left(\frac{5}{9}+\frac{e^2}{2}\right)M_{\saturn}\frac{R^5}{a^3}}, \label{eq:J1} 
\end{eqnarray}
and this results in 

\begin{equation}
\label{eq:smallsigma}
\sigma(t) = \frac{\kappa_1}{\omega_0^2-(n-\dot{\varpi})^2}\sin\mathcal{M},
\end{equation}
after damping of the proper oscillations, their frequency being $\omega_0$. Here the semi-diurnal oscillations have been dropped. The numerical application
shows that their amplitude is negligible.

\par The Tab.~\ref{tab:comparison} proposes a comparison between this study and two previous ones, by \citet{vbt2013} and \citet{rrc2014}. These two studies
aimed at modeling the librations of a satellite composed of a viscoelastic crust coating a global ocean, itself enshrouding a rigid core. Hence, they consider
pressure and gravitational couplings between the different layers, that I do not have here. In considering only the outer shell in their studies, I managed
to get equations like my Eq.~(\ref{eq:smalllibra}), from \citep[Eq.~45]{vbt2013} and \citep[App.~A]{rrc2014}, with the help of \citep{r2014}.

\placetable{tab:comparison}

\par In those two studies, $A$, $B$ and $C$ stand for the mean values of the diagonal elements of the tensor of inertia $\mathcal{I}$, i.e. $I_{11}$, $I_{22}$ and $I_{33}$.
Moreover, the triaxiality is supposed to be entirely due to the rotational and tidal deformations, i.e. there is no frozen component in $(I_{22}-I_{11})$, and the second
degree in the eccentricity has been neglected. In \citep{vbt2013}, $k_2$ stands for $k_2(\nu_1)$, while in \citep{rrc2014}, the topographical Love number $h_2$ has been used. 

\par My formula for $\omega_0$ is consistent with the one of \citet{rrc2014}, while \citet{vbt2013} have a signature of the tidal deformation at the diurnal frequency.
However, these two previous studies are consistent with each other for the coefficient $\kappa_1$ for $h_2=5k_2/3$.

\par A difference between these studies and mine is that they assume the resonant argument $\sigma$ to be fixed to $0$ in the expression of the tensor of inertia 
(Eq.~\ref{eq:I11ortid} to \ref{eq:I22ortid}), while it is a variable of the problem.  If we set $\sigma = 0$ in the tensor of inertia, then the gravitational torque 
$\Gamma$ defined by the Eq.~\ref{eq:Gammaz2} becomes, after averaging over the mean anomaly $\mathcal{M}$:

\begin{equation}
\label{eq:wronggamma}
\Gamma = 3\frac{\mathcal{G}M_{\saturn}}{r^3}\left(I_{22}^{(f)}-I_{11}^{(f)}\right)\sin2\sigma
\end{equation}
i.e. only the frozen component remains on average. If we had only an elastic inertia, then the average torque would be 

\begin{equation}
\label{eq:wronggamma2}
\Gamma = 3\frac{\mathcal{G}M_{\saturn}}{r^3}\left(k_fM_{\saturn}\frac{R^5}{a^3}\cos2\sigma\sin2\sigma-k_fM_{\saturn}\frac{R^5}{a^3}\cos2\sigma\sin2\sigma\right) = 0,
\end{equation}
that comes from

\begin{eqnarray}
xy & = & \frac{\sin2\sigma}{2}+\mathcal{O}(e), \label{eq:xywrong} \\
(I_{22}-I_{11})^{(e)} & = & k_fM_{\saturn}\frac{R^5}{a^3}\cos2\sigma+\mathcal{O}(e), \label{eq:BmAwrong} \\
x^2-y^2 & = & \cos2\sigma+\mathcal{O}(e), \label{eq:x2y2wrong} \\
I_{12}^{(e)} & = & -\frac{k_f}{2}M_{\saturn}\frac{R^5}{a^3}\sin2\sigma+\mathcal{O}(e). \label{eq:I12wrong}
\end{eqnarray}
So, a frozen component in $(I_{22}-I_{11})$ is needed to get a non-null mean gravitational torque.

\section{Modeling the tides\label{sec:tides}}

\subsection{The Maxwell model}

\par The classical Maxwell model \citep[e.g.]{k2008} gives a pretty good estimation of the frequency-dependency of $k_2$. It depends on one parameter, 
the Maxwell time $\tau_M = \eta/\mu$, $\eta$ being the viscosity and $\mu$ the rigidity. The complex Love number $k_2^*$ reads

\begin{equation}
\label{eq:k2star}
k_2^* = k_f\frac{J^*(\nu)}{J^*(\nu)+A_2/\mu}
\end{equation}
with

\begin{equation}
\label{eq:A2}
A_2 = \frac{19}{2}\frac{\mu}{\rho gR},
\end{equation}
and $J^*$ is the complex compliance defined as:

\begin{equation}
\label{eq:Jstar}
J^* = \frac{\nu\tau_M-\imath}{\mu\nu\tau_M},
\end{equation}
$\nu$ being the tidal frequency, and $g$ the surface gravity of the body. $k_2$ is the real part of $k_2^*$, i.e. 

\begin{equation}
\label{eq:rek2}
\mathfrak{R}(k_2^*) = k_f\frac{1+(1+A_2)\nu^2\tau_M^2}{1+(1+A_2)^2\nu^2\tau_M^2}.
\end{equation}

\subsection{An improvement at high frequencies}

\par At high frequencies, the Maxwell model lacks accuracy because it does not render the fact that anelasticity dominates viscoelasticity, i.e. the response of 
the material is not instantaneous anymore. The Andrade model is therefore more appropriate. This is the reason why \citet{e2012a} proposed the so-called 
Andrade-Maxwell model, that corresponds to the Andrade model at high frequencies and to the Maxwell model at lower frequencies. Its complex compliance reads:

\begin{equation}
\label{eq:Efroimskycompliance}
\bar{J}(\nu) = \left(1+\left(\imath\nu\tau_A\right)^{-N}\Gamma(1+N)-\imath\left(\nu\tau_M\right)^{-1}\right)/\mu,
\end{equation}
$\Gamma$ being the classical $\Gamma$ function defined as

\begin{equation}
\label{eq:Gamma}
\Gamma(1+N) = \int_0^{+\infty} z^N e^{-z}\,\textrm{d}z.
\end{equation}

\par We can see that this model depends on 3 tidal parameters, which are the Maxwell time $\tau_M$, an Andrade time $\tau_A$ that has been introduced by 
\citet[Eq.~78]{e2012a}, and an Andrade parameter $N$. The Andrade time should be equal to the Maxwell time to have a continuous transition between viscoelasticy and 
anelasticity, i.e. $\tau_A = \tau_M$. $N$ is usually assumed to lie between 0.1 and 0.5, I will take the classical value $N=0.3$. The resulting expression for the Andrade-Maxwell model is

\begin{equation}
\label{eq:Rek2Efroimsky}
\mathfrak{R}(k_2^*) = k_f\frac{\mathcal{A}^2+\mathcal{A}A_2+\mathcal{B}^2}{\mathcal{A}^2+2\mathcal{A}A_2+A_2^2+\mathcal{B}^2}
\end{equation}
with

\begin{eqnarray}
\mathcal{A} & = & 1+\left(\nu\tau_A\right)^{-N}\cos\left(\frac{N\pi}{2}\right)\Gamma(1+N), \label{eq:Aronde} \\
\mathcal{B} & = & \left(\nu\tau_M\right)^{-1}+\left(\nu\tau_A\right)^{-N}\sin\left(\frac{N\pi}{2}\right)\Gamma(1+N). \label{eq:Bronde}
\end{eqnarray}

\par The Andrade parameter $\tau_A$ is an order of magnitude of the period above which the excitation will generate the Andrade creep, responsible for anelasticity. Setting $\tau_A$ to the 
infinity renders the Maxwell rheology. 

\par
\begin{figure}[ht]
\centering
\begin{tabular}{cc}
\includegraphics[width=8cm,height=5.5cm]{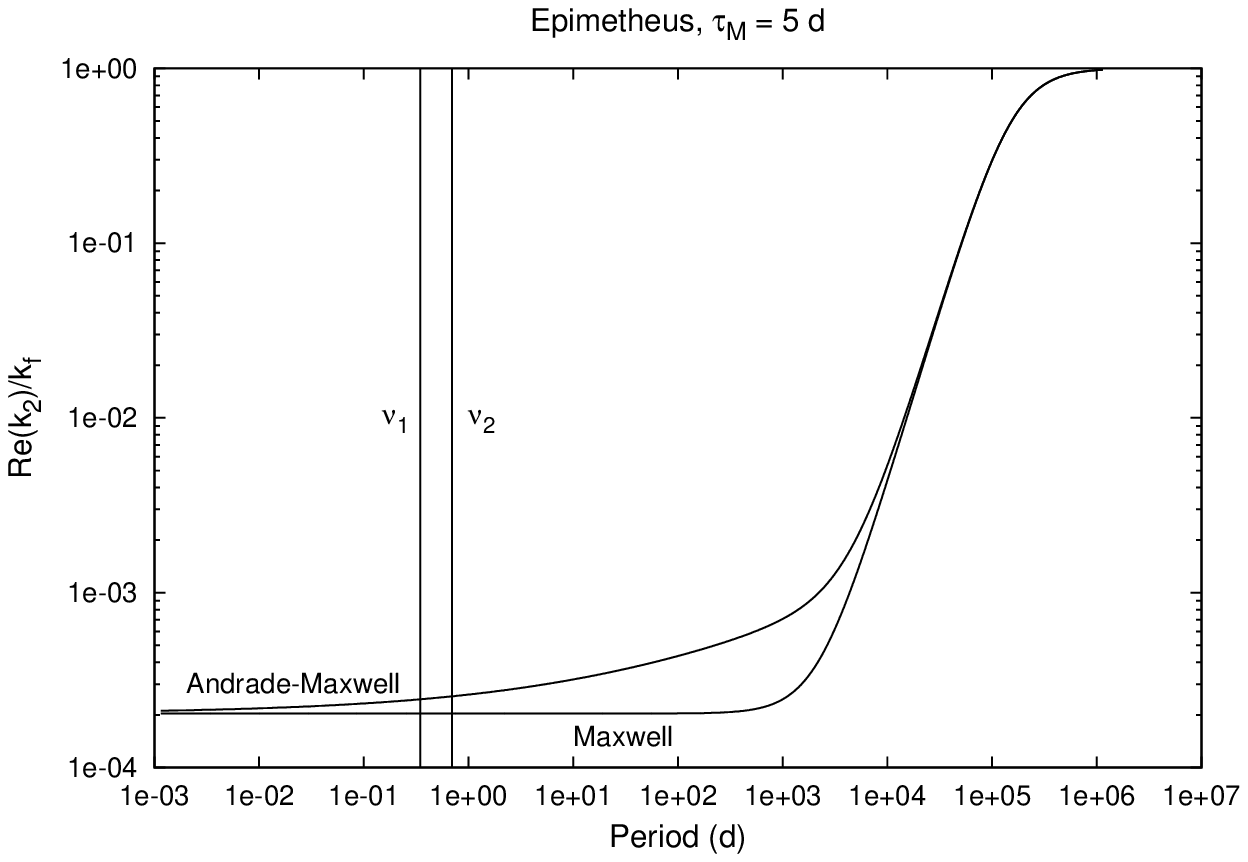} & \includegraphics[width=8cm,height=5.5cm]{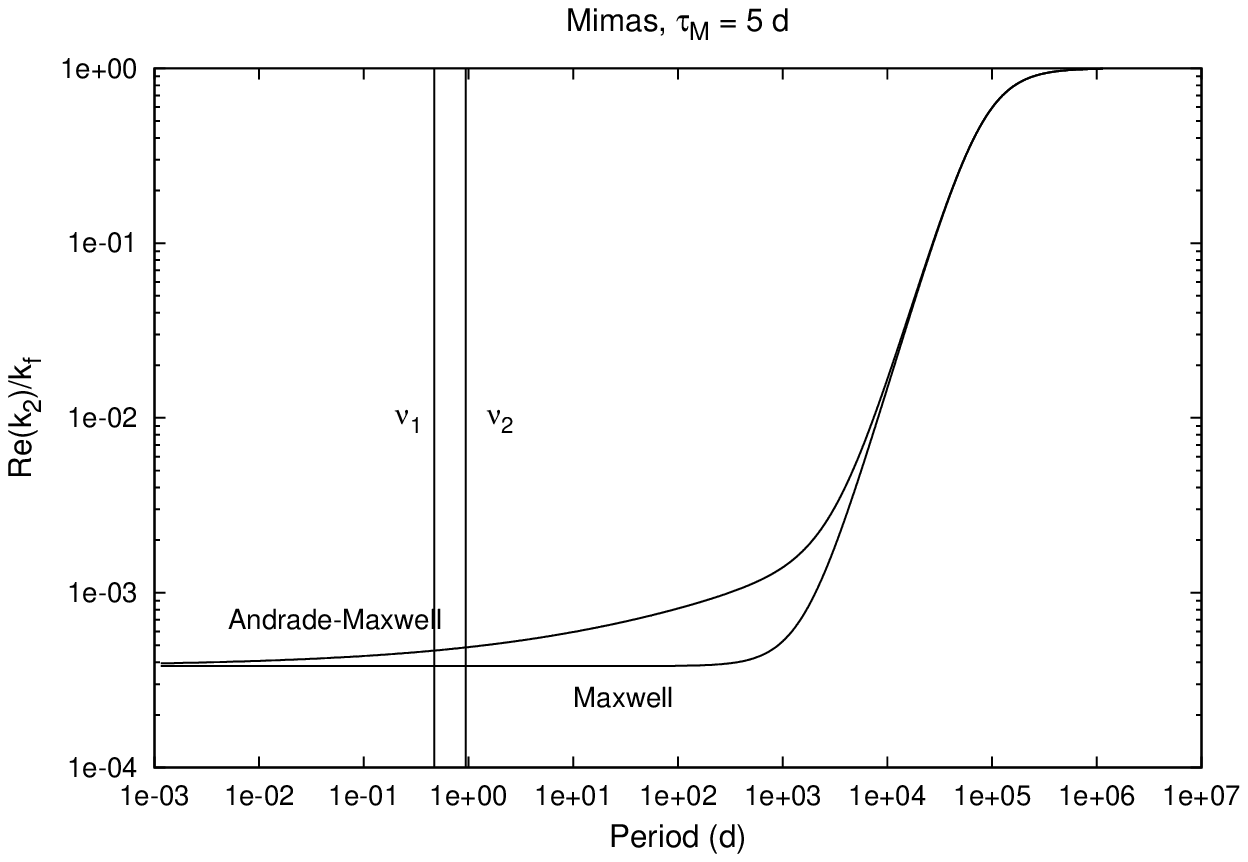}
\end{tabular}
\caption[Elastic tides]{Elastic tides for Epimetheus and Mimas. The Andrade time is set equal to the Maxwell time, and the Andrade parameter $N$ to $0.3$.
The two vertical lines represent the diurnal and semi-diurnal periods. We can see that there are few differences between the two models. \label{fig:rek2}}
\end{figure}

 The Fig.~\ref{fig:rek2} illustrates the elastic tides given by these two models, for Mimas and Epimetheus, in using the physical parameters given in Tab.~\ref{tab:param}.

\placefigure{fig:rek2}

\section{Application to Epimetheus and Mimas\label{sec:applications}}

\par Diurnal librations have been measured for Epimetheus and Mimas, thanks to Cassini data \citep{ttb2009,trlcrrn2014}.  These two bodies are a priori assumed
to be solid bodies, which legitimates the use of this model to explain their librations.

\subsection{Methodology}

\par A numerical integration of the Eq.(\ref{eq:biglibra}) is performed, with the 10th order Adams-Bashforth-Moulton predictor-corrector scheme. A frequency analysis is 
then made to decompose the variable $\sigma(t)$ under a quasi-periodic formulation. This way, the proper oscillations and the forced ones are clearly identified. The frequency 
analysis algorithm is based on Laskar's original idea \citep{l1999,l2005}, has been adapted for the rotational dynamics by \citet{nlv2008}, and used many times since. The 
initial conditions of the numerical integration are given by the analytical solution at $t=0$, i.e. the Eq.(\ref{eq:smallsigma}) and its time-derivative, while the parameters 
are gathered in Tab.~\ref{tab:param}.

\par For each of these satellites, several simulations are run, which differ by the fraction of non-hydrostaticity. This starts from the formula giving the mean value of $I_{22}-I_{11}$:

\begin{equation}
\label{eq:I2211}
<I_{22}-I_{11}> \approx (I_{22}-I_{11})^{(f)}+k_f\frac{M_{\saturn}R^5}{a^3}\cos 2<\sigma>,
\end{equation}
and then, two approaches are considered:

\begin{enumerate}

\item either the secular Love number $k_f$ is set to be constant. I first set it to $1.5$ before discussing the implications of a smaller $k_f$, 
which could be more physically relevant. I define the parameter $\alpha$ such that

\begin{equation}
\label{eq:nonhydroa}
(I_{22}-I_{11})^{(f)} = \alpha k_f\frac{M_{\saturn}R^5}{a^3}.
\end{equation}

This approach suggests that the body is composed of the superimposition of 2 mass distributions, an elastic and a frozen ones. The 
elastic contribution is fully consistent with the hydrostatic theory, and the frozen contribution supplements the elastic one, as an excess of
triaxiality. This excess represents a fraction $\alpha$ of the elastic contribution,

\item or the mean value of $(I_{22}-I_{11})$ is considered to be constant and determined from the shape of the satellite in assuming
a constant density, i.e.

\begin{equation}
\label{eq:I2211shape}
<I_{22}-I_{11}> = \frac{M}{5}\left(\mathfrak{a}^2-\mathfrak{b}^2\right),
\end{equation}
which implies

\begin{equation}
\label{eq:C22shape}
<C_{22}> = \frac{\mathfrak{a}^2-\mathfrak{b}^2}{20 R^2}.
\end{equation}

I define the parameter $\beta$ such that

\begin{equation}
\label{eq:nonhydrob}
(I_{22}-I_{11})^{(f)} = \beta\frac{M}{5}\left(\mathfrak{a}^2-\mathfrak{b}^2\right),
\end{equation}
and $k_f$ is obtained from the formula (\ref{eq:I2211}). This would mean that the mean inertia of the body is consistent with the hydrostatic theory,
but part of this inertia is frozen, the remaining part being still elastic.

\end{enumerate}

\placetable{tab:param}

\subsection{Epimetheus}

\par For Epimetheus, a physical libration of $5.9\pm1.2^{\circ}$ has been measured by \citet{ttb2009}. Such a large number is due to the triaxiality of the satellite,
which makes the frequency of the proper oscillations $\omega_0$ close to the orbital one, resulting in a large amplitude of response \citep[Fig.~6]{n2010}. 

\begin{figure}[ht]
\centering
\begin{tabular}{cc}
\includegraphics[width=8cm,height=5.5cm]{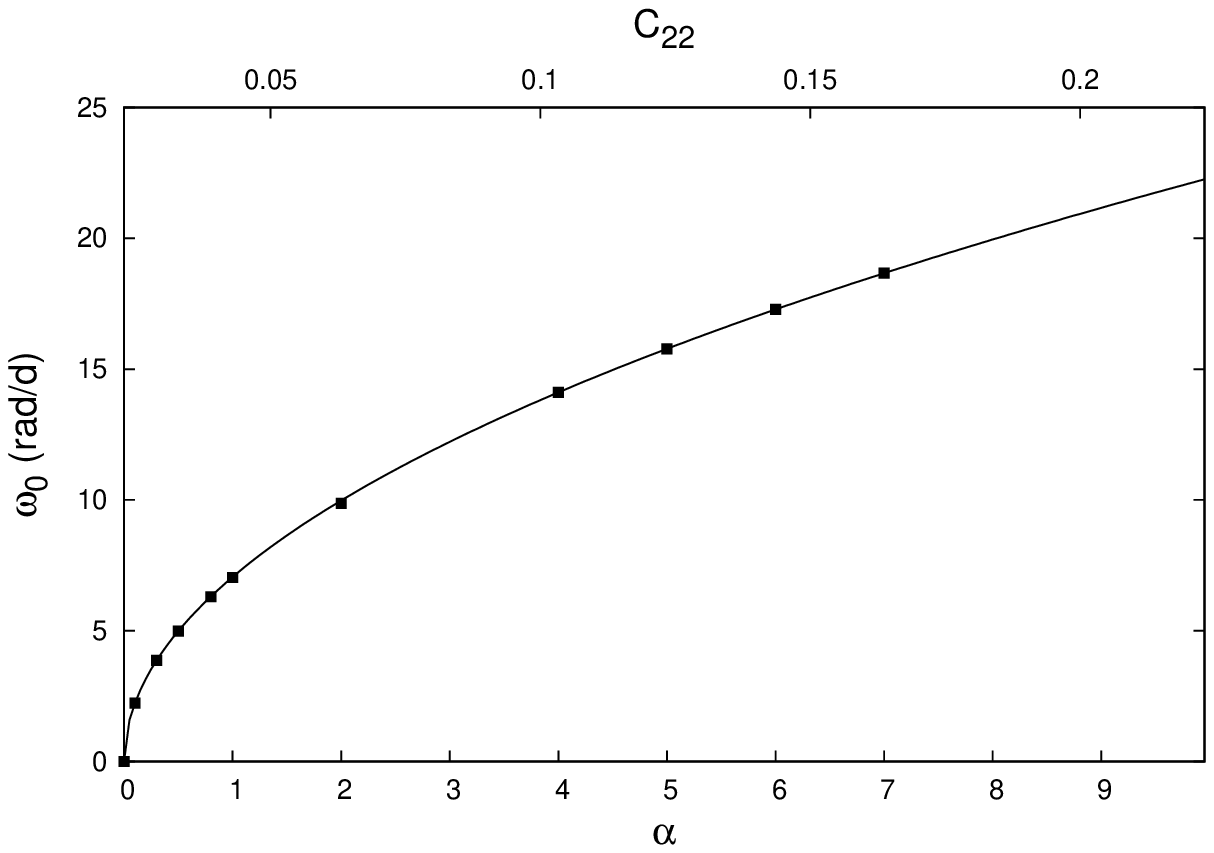}  & \includegraphics[width=8cm,height=5.5cm]{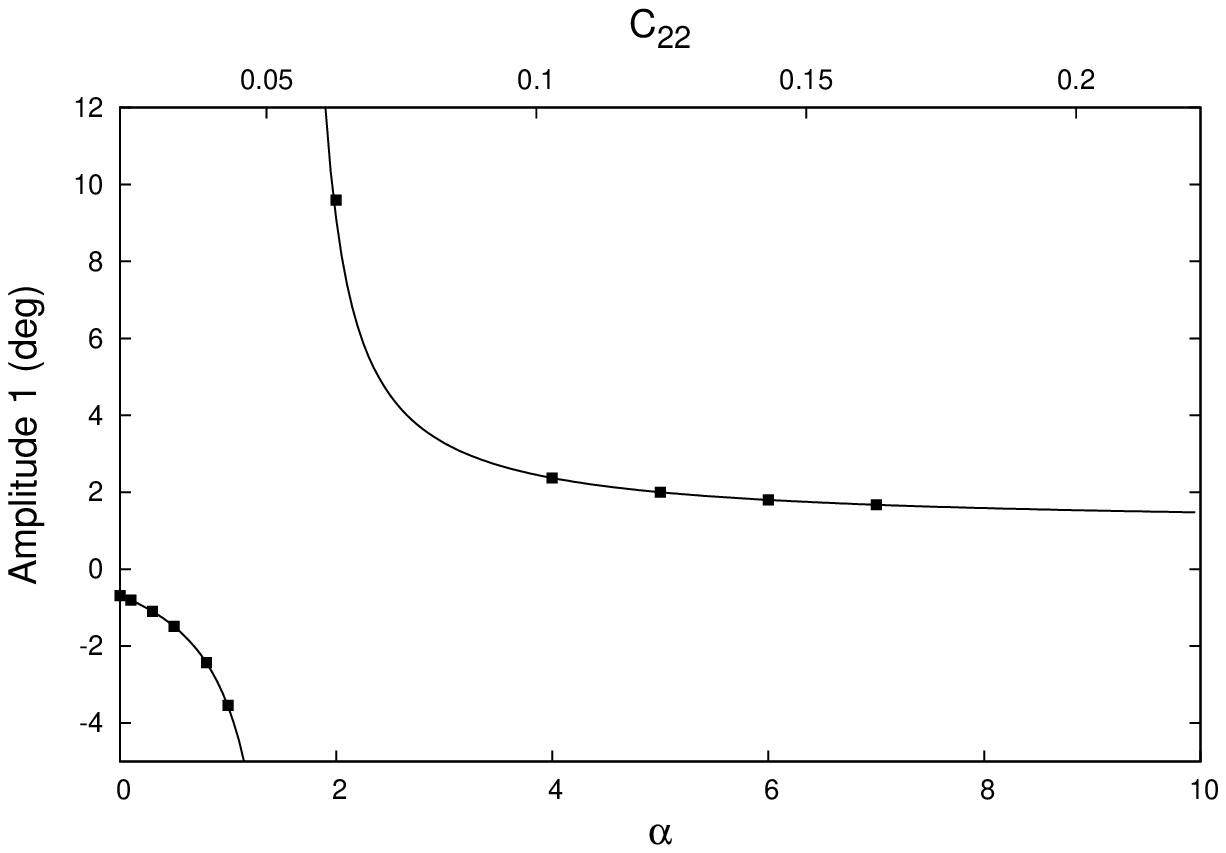}
\end{tabular}
\caption[Rotational quantities for Epimetheus, in considering an excess of triaxiality]{Rotational quantities for Epimetheus, in considering an excess of triaxiality, for $k_f=1.5$. The lines come from the analytical 
formulae (\ref{eq:I2211}, \ref{eq:omega02} \& \ref{eq:smallsigma}), while the squares are the results from numerical simulations. The \emph{Amplitude 1} is the amplitude of response
at the diurnal frequency $n-\dot{\varpi}$.\label{fig:epimkf}}
\end{figure}

\placefigure{fig:epimkf}

\par The simulations with an excess of triaxiality result in a larger $C_{22}$ (Fig.~\ref{fig:epimkf}) than anticipated by the measured shape, but permit
to reach the measured amplitude for $\alpha=1.21_{-0.09}^{+0.06}$. 

\begin{figure}[ht]
\centering
\includegraphics[width=12cm,height=8.3cm]{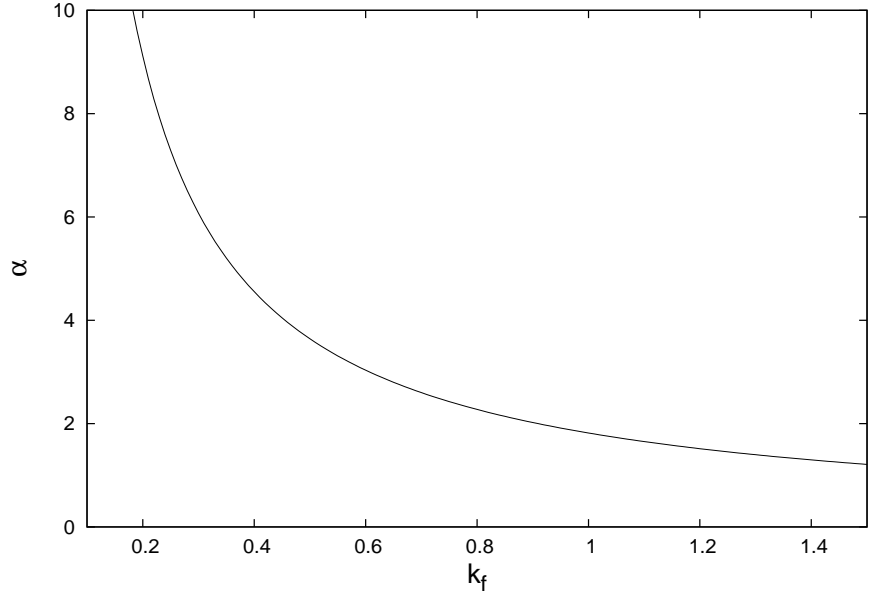}
\caption{Excess of triaxiality for Epimetheus, for different $k_f$.\label{fig:epimdiffkf}}
\end{figure}

\placefigure{fig:epimdiffkf}

$k_f=1.5$ for the elastic part should be seen as an incomplete model, in the sense that no rigidity is considered. The Fig.~\ref{fig:epimdiffkf} displays $\alpha$
for smaller values of $k_f$. In particular, we have $\alpha=1.82_{-0.11}^{+0.08}$ for $k_f=1$.

These numbers should be balanced by the uncertainties of a few kilometers on the radii $\mathfrak{a}$ and $\mathfrak{b}$, and some departures from an actual triaxial shape. We can also 
see that the analytical formulae for $\omega_0$ (Eq.~\ref{eq:omega02}) and the amplitude 1 (Eq.~\ref{eq:smallsigma}) are validated by the numerical simulations.

\begin{figure}[ht]
\centering
\begin{tabular}{cc}
\includegraphics[width=8cm,height=5.5cm]{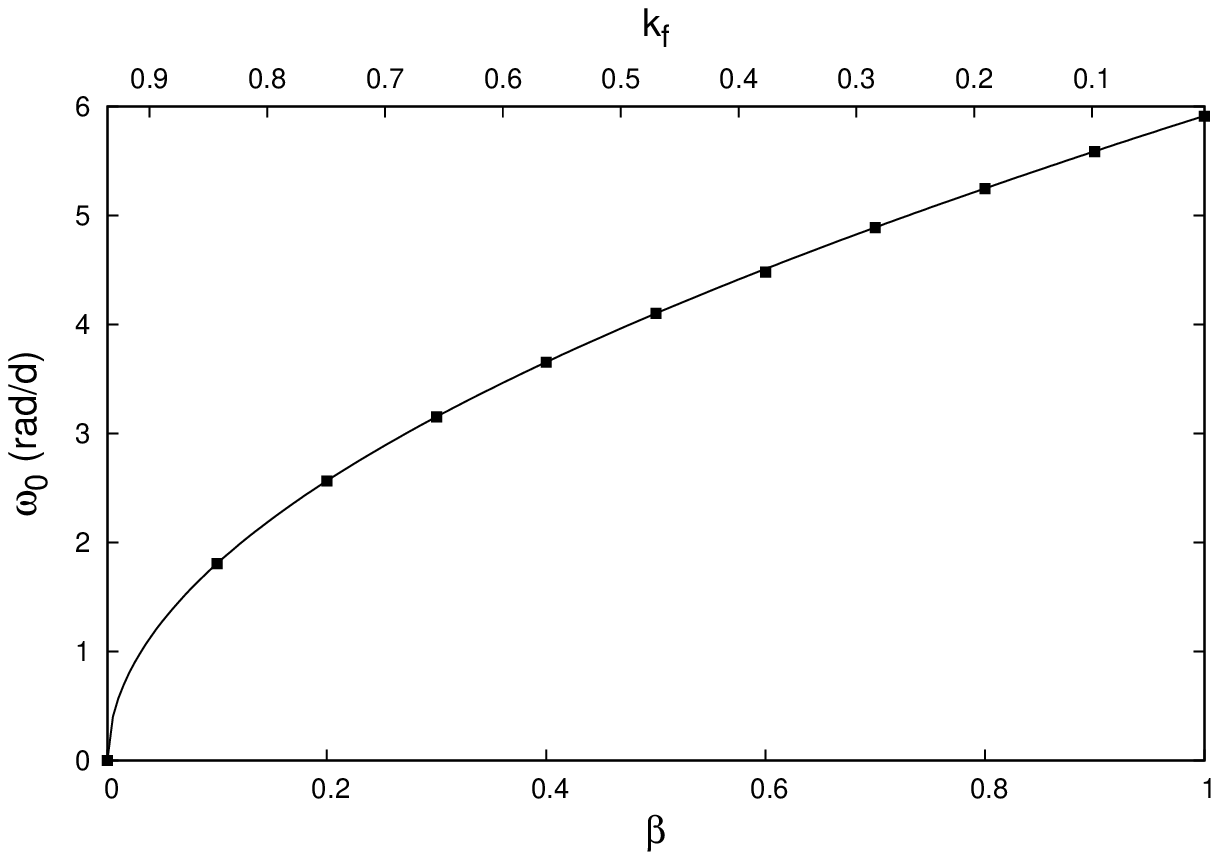} & \includegraphics[width=8cm,height=5.5cm]{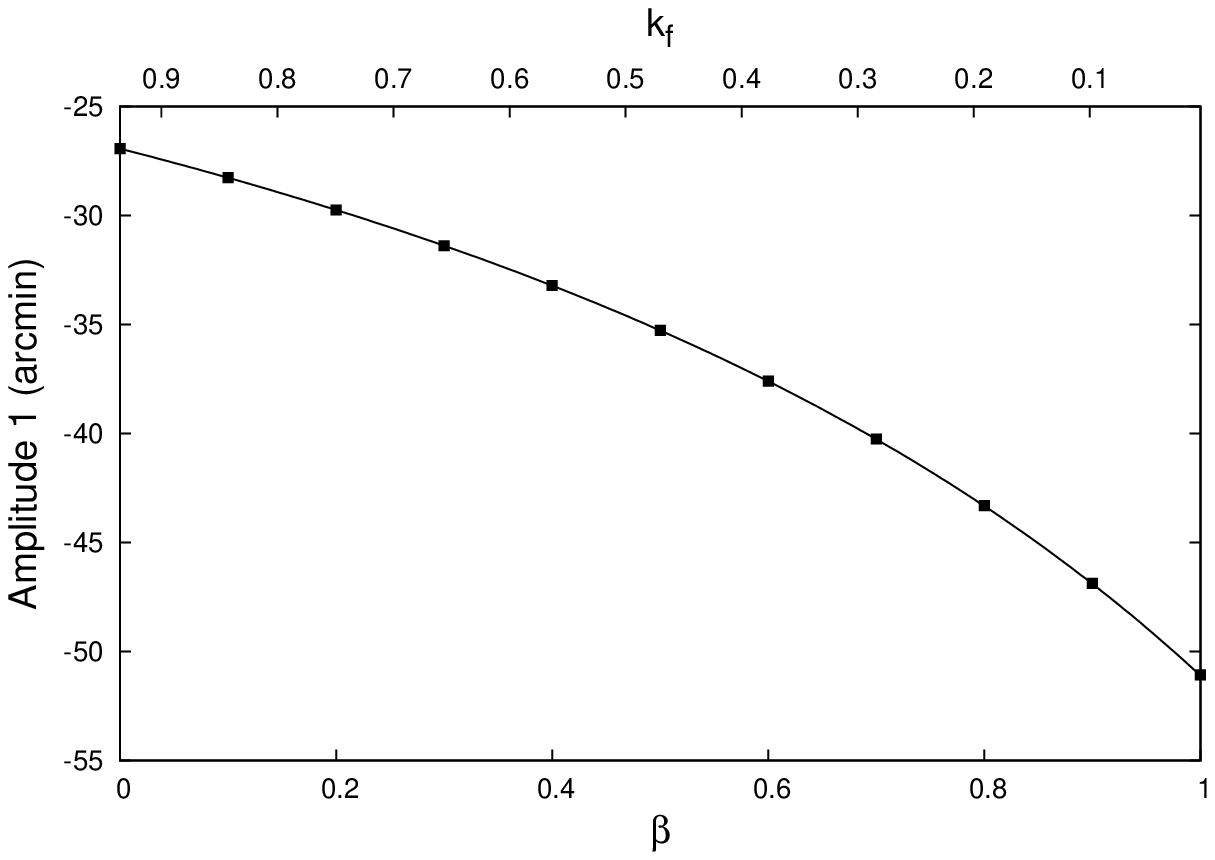}
\end{tabular}
\caption[Rotational quantities for Epimetheus, for $C_{22}=1.426\times10^{-2}$]{Rotational quantities for Epimetheus, for $C_{22}=1.426\times10^{-2}$, as
suggested by the measured shape. The lines come from the analytical formulae, while the squares are the results of numerical simulations.\label{fig:epimetheusC22}}
\end{figure}

\placefigure{fig:epimetheusC22}

\par The simulations assuming a constant triaxiality for Epimetheus (Fig.~\ref{fig:epimetheusC22}) still validate the analytical formulae for $\omega_0$ and 
the diurnal amplitude. Moreover, they do not result in a large enough amplitude to match the observed libration.

\par This study assumes that Epimetheus is triaxial. Actually the visual aspect of Epimetheus suggests significant departure from the ellipsoid. To the best of
my knowledge, no higher order shape model has been published, this is why I assume the triaxial ellipsoid as a good enough model. Of course, the departures to 
the triaxiality should contribute to the librations.

\subsection{Mimas}

\par \citet{trlcrrn2014} have measured from Cassini data a surprisingly high diurnal amplitude for Mimas of $50.3\pm1.0$ arcmin, while a theoretical study
by \citet{nkr2011} predicted half this number. That study assumed Mimas to be a rigid body made of a mixture of ice and silicates, with a radial gradient of porosity.
It was modeled as two homogeneous layers, the inner one being denser than the outer one. \citet{trlcrrn2014} tested 5 plausible interiors to explain their
measurements, and kept only two: Mimas had either a global ocean beneath an icy crust, or a highly elongated core of pretty heavy elements. This last explanation 
would be consistent with a model of formation of the Saturnian satellites proposed by \citet{cccldktmls2011}. In that scenario, the satellites would have been 
formed as droplets from the rings before migrating outward to their present location. The highly triaxial shape of the core would be the signature of a former 
more rapid rotation, itself meaning that the shape froze when Mimas was closer to Saturn than it is now. It translates into an excess of triaxiality.

\begin{figure}[ht]
\centering
\begin{tabular}{cc}
\includegraphics[width=8cm,height=5.5cm]{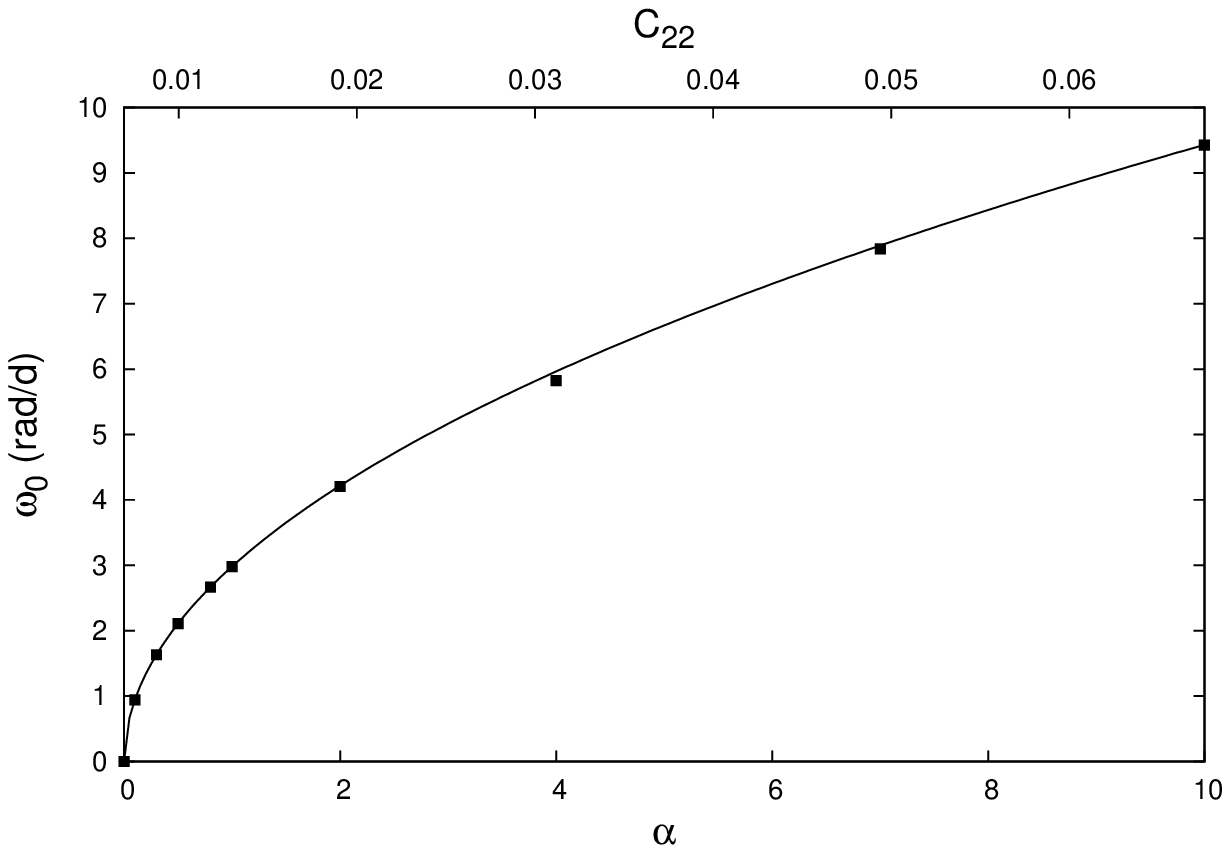} & \includegraphics[width=8cm,height=5.5cm]{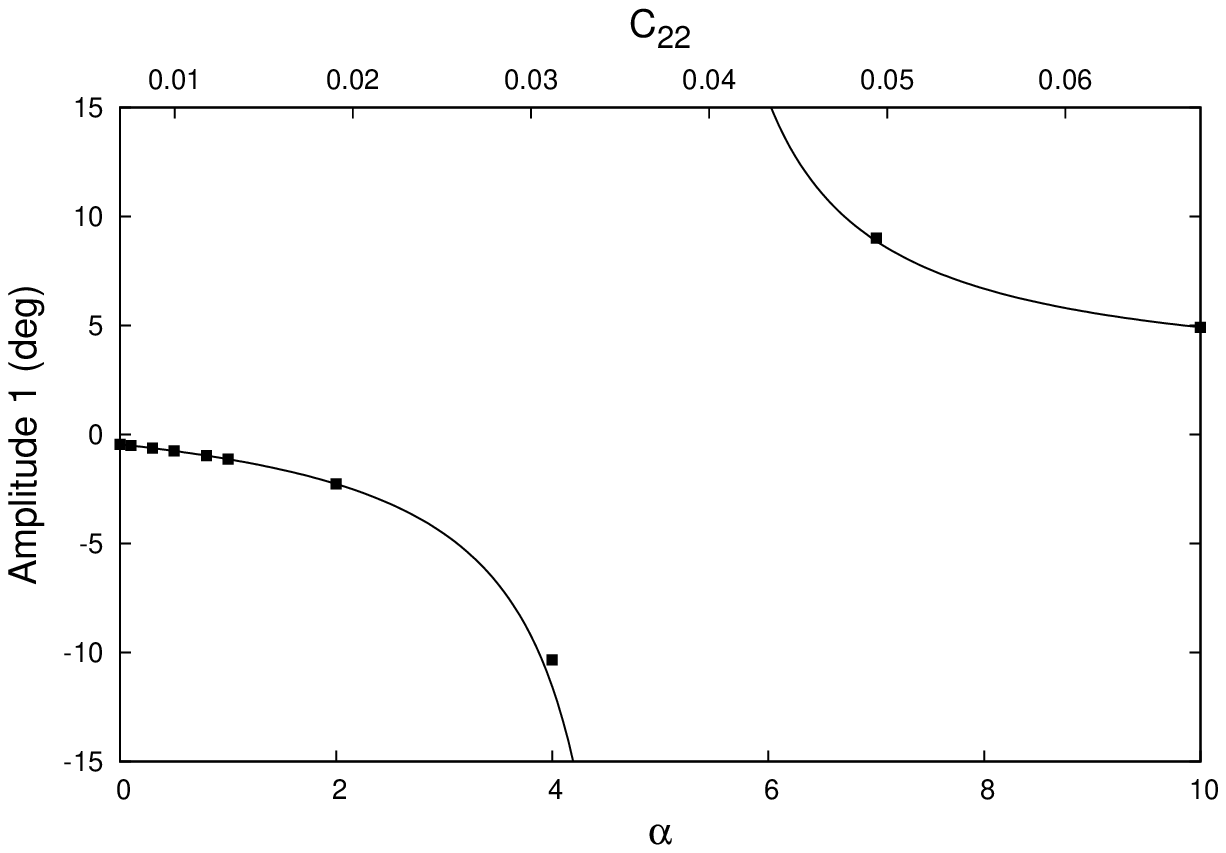}
\end{tabular}
\caption[Rotational quantities for Mimas, with an excess of triaxiality]{Rotational quantities for Mimas, with an excess of triaxiality, for $k_f=1.5$.\label{fig:mimaskf}}
\end{figure}

\placefigure{fig:mimaskf}

\begin{figure}[ht]
\centering
\includegraphics[width=12cm,height=8.3cm]{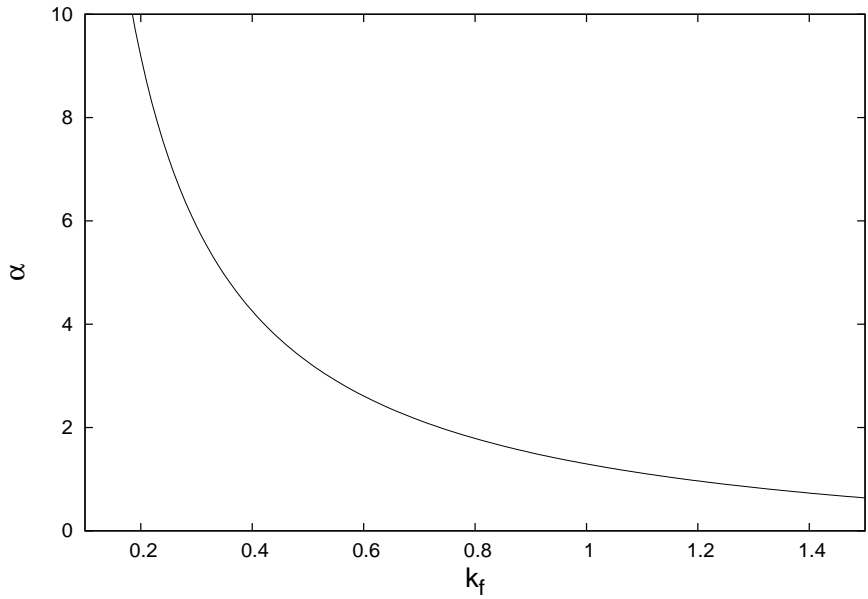}
\caption{Excess of triaxiality for Mimas, for different $k_f$.\label{fig:mimasdiffkf}}
\end{figure}

\placefigure{fig:mimasdiffkf}

\par
\begin{figure}[ht]
\centering
\begin{tabular}{cc}
\includegraphics[width=8cm,height=5.5cm]{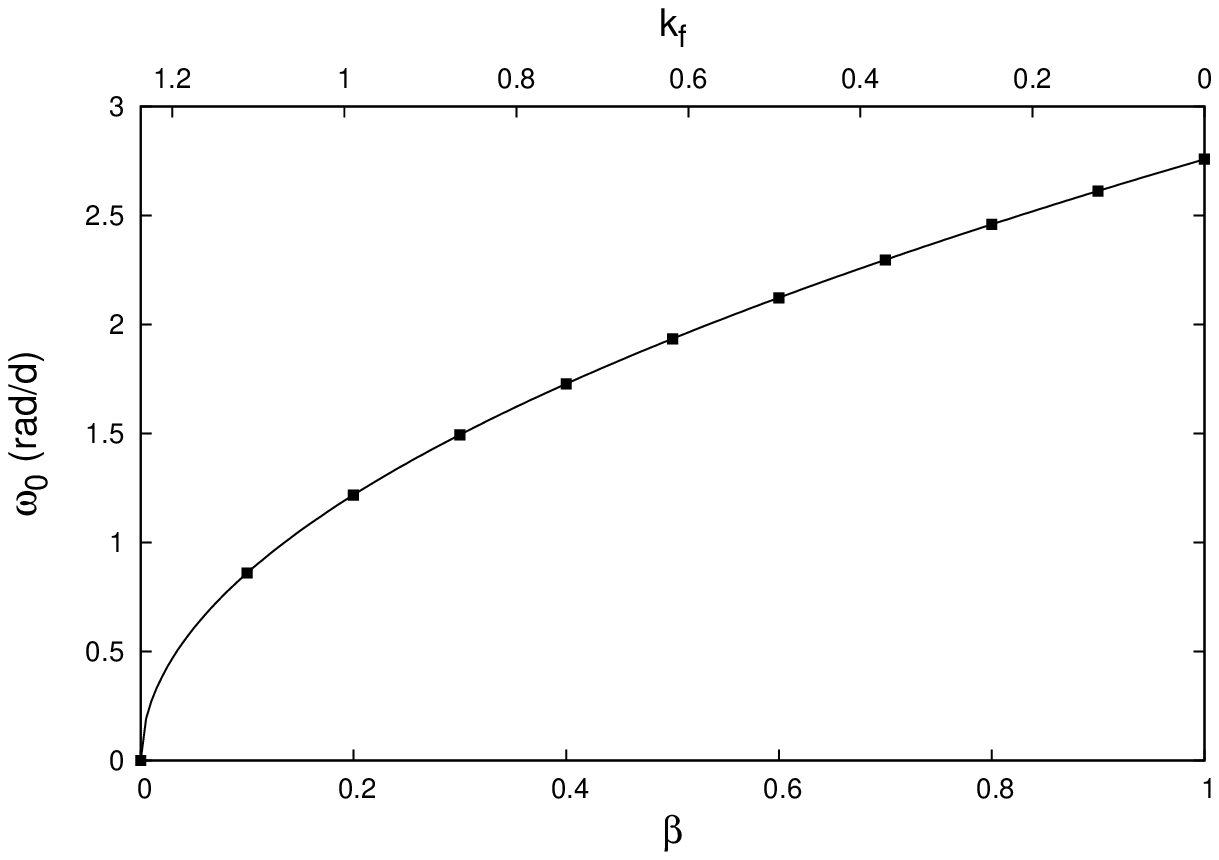} & \includegraphics[width=8cm,height=5.5cm]{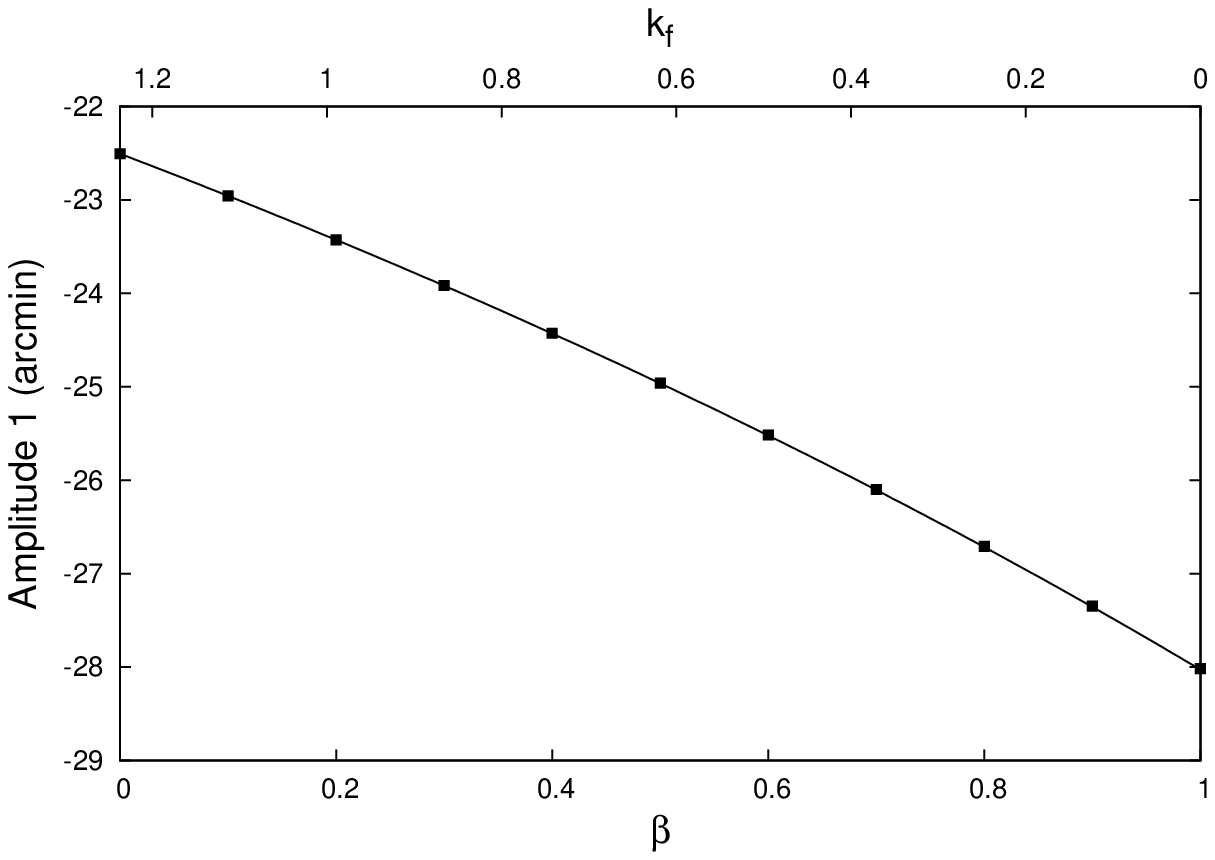}
\end{tabular}
\caption[Rotational quantities for Mimas, for $C_{22}=5.606\times10^{-3}$]{Rotational quantities for Mimas, for $C_{22}=5.606\times10^{-3}$, deduced from the measured shape.\label{fig:mimasC22}}
\end{figure}

 The two figures \ref{fig:mimaskf} and \ref{fig:mimasC22} still validate the analytical formulae for $\omega_0$ and the diurnal amplitude. 
Moreover, they permit the measured amplitude for $\alpha = 0.6225\pm0.0235$ and $k_f=1.5$, while a constant $C_{22}$ prohibits it. 
$k_f=1$ would imply $\alpha=1.2954\pm0.0332$ (Fig.~\ref{fig:mimasdiffkf}).

\placefigure{fig:mimasC22}

\section{Influence of the creep\label{sec:dissipative}}

\par I now introduce the imaginary part of the Love number, that is often denoted as  $\mathfrak{I}(k_2^*) = -k_2/Q$, where $Q$ is a dissipation function. 
This imaginary part introduces a dissipation as a lag between the gravitational excitation of the parent planet and the 
response of the satellite. This lag depends on the tidal frequency, this is why this effect will appear as $\left(\frac{k_2}{Q}\right)(\nu)$.

\subsection{Equations of the problem}

\par As shown for instance in \citep{wb2015}, the dissipative tide can be introduced in replacing in the formulae (\ref{eq:I11ortid}) to (\ref{eq:I33ortid}) 
$k_2\cos$ by $k_2\cos+k_2/Q\sin$ and $k_2\sin$ by $k_2\sin-k_2/Q\cos$. This geometrically means that the parent planet raises two bulges at the surface of 
the satellite, i.e. a bulge which is aligned with the direction of the planet, and which is responsible for the elastic tide, and a bulge in quadrature, which is responsible 
for the dissipative tide \citep[e.g.]{z1966,e2012b}. The elastic inertia $\mathfrak{I}^{(e)}$ is replaced by a viscoelastic one $\mathfrak{I}^{(v)}$. This yields:

\begin{eqnarray}
I_{11}^{(v)} & = & \frac{M_{\saturn}R^5}{a^3}\left(k_2(\nu_0)\left(-\frac{5}{18}-\frac{e^2}{4}-\frac{1}{2}\left(1-\frac{5}{2}e^2\right)\cos 2\sigma\right)\right. \nonumber \\
& & \left.+k_2(\nu_1)\frac{e}{4}\left(\cos(\mathcal{M}+2\sigma)-2\cos(\mathcal{M})-7\cos(\mathcal{M}-2\sigma)\right)\right. \nonumber \\
& & \left.+\left(\frac{k_2}{Q}\right)(\nu_1)\frac{e}{4}\left(\sin(\mathcal{M}+2\sigma)-2\sin(\mathcal{M})-7\sin(\mathcal{M}-2\sigma)\right)\right. \nonumber \\
& & \left.-k_2(\nu_2)\frac{e^2}{4}\left(3\cos(2\mathcal{M})+17\cos(2\mathcal{M}-2\sigma)\right)\right. \nonumber \\
& & \left.-\left(\frac{k_2}{Q}\right)(\nu_2)\frac{e^2}{4}\left(3\sin(2\mathcal{M})+17\sin(2\mathcal{M}-2\sigma)\right)\right), \label{eq:I11ortiddis} \\
I_{12}^{(v)} & = & \frac{M_{\saturn}R^5}{2a^3}\left(k_2(\nu_0)\left(1-\frac{5}{2}e^2\right)\sin 2\sigma-k_2(\nu_1)\frac{e}{2}\left(\sin(\mathcal{M}+2\sigma)+7\sin(\mathcal{M}-2\sigma)\right)\right. \nonumber \\
& & \left.+\left(\frac{k_2}{Q}\right)(\nu_1)\frac{e}{2}\left(\cos(\mathcal{M}+2\sigma)+7\cos(\mathcal{M}-2\sigma)\right)\right. \nonumber \\
& & \left.-\frac{17}{2}e^2\left(k_2(\nu_2)\sin(2\mathcal{M}-2\sigma)-\left(\frac{k_2}{Q}\right)(\nu_2)\cos(2\mathcal{M}-2\sigma)\right)\right), \label{eq:I12ortiddis} \\
I_{22}^{(v)} & = & \frac{M_{\saturn}R^5}{a^3}\left(k_2(\nu_0)\left(-\frac{5}{18}-\frac{e^2}{4}+\frac{1}{2}\left(1-\frac{5}{2}e^2\right)\cos 2\sigma\right)\right. \nonumber \\
& & \left.+\frac{e}{4}k_2(\nu_1)\left(7\cos(\mathcal{M}-2\sigma)-\cos(\mathcal{M}+2\sigma)-2\cos(\mathcal{M})\right)\right. \nonumber \\
& & \left.+\frac{e}{4}\left(\frac{k_2}{Q}\right)(\nu_1)\left(7\sin(\mathcal{M}-2\sigma)-\sin(\mathcal{M}+2\sigma)-2\sin(\mathcal{M})\right)\right. \nonumber \\
& & \left.+\frac{e^2}{4}k_2(\nu_2)\left(17\cos(2\mathcal{M}-2\sigma)-3\cos(2\mathcal{M})\right)\right. \\
& & \left.+\frac{e^2}{4}\left(\frac{k_2}{Q}\right)(\nu_2)\left(17\sin(2\mathcal{M}-2\sigma)-3\sin(2\mathcal{M})\right)\right), \label{eq:I22ortiddis} \\
I_{33}^{(v)} & = & \frac{M_{\saturn}R^5}{a^3}\left(k_2(\nu_0)\left(\frac{5}{9}+\frac{1}{2}e^2\right)+e\left(k_2(\nu_1)\cos(\mathcal{M})+\left(\frac{k_2}{Q}\right)(\nu_1)\sin(\mathcal{M})\right)\right. \nonumber \\
& & +\left.\frac{3}{2}e^2\left(k_2(\nu_2)\cos(2\mathcal{M})+\left(\frac{k_2}{Q}\right)(\nu_2)\sin(2\mathcal{M})\right)\right). \label{eq:I33ortiddis} 
\end{eqnarray}

\par If we consider the Maxwell rheology, then the quantity $k_2/Q = -\mathfrak{I}(k_2^*)$ is obtained from the formula (\ref{eq:k2star}), i.e.

\begin{equation}
\label{eq:imk2}
\left(\frac{k_2}{Q}\right)(\nu) = k_f\frac{A_2\nu\tau_M}{1+(1+A_2)^2\nu^2\tau_M^2}.
\end{equation}
It is in particular expected from any rheology that $k_2/Q$ is null at the zero frequency, i.e. no dissipation occurs for a constant excitation.

\par The gravitational torque $\Gamma$ (Eq.~\ref{eq:Gammadev}) then becomes 

\begin{eqnarray}
\Gamma & = & 18n^2e^2\frac{M_{\saturn}R^5}{a^3}\left(\frac{k_2}{Q}\right)(\nu_1)+\left(-\frac{3}{2}+\frac{15}{4}e^2\right)\left(I_{22}-I_{11}\right)^{(f)}n^2\sin2\sigma \nonumber \\
& & +\frac{3}{4}\left(I_{22}-I_{11}\right)^{(f)}en^2\left(\sin(\mathcal{M}+2\sigma)+7\sin(\mathcal{M}-2\sigma)\right) \nonumber \\
& & +\frac{51}{4}\left(I_{22}-I_{11}\right)^{(f)}e^2n^2\sin(2\mathcal{M}-2\sigma)+6en^2\frac{M_{\saturn}R^5}{a^3}\left(k_f-k_2(\nu_1)\right)\sin\mathcal{M} \nonumber \\
& & +6en^2\frac{M_{\saturn}R^5}{a^3}\left(\frac{k_2}{Q}\right)(\nu_1)\cos\mathcal{M} \nonumber \\
& & +\frac{51}{4}n^2e^2\frac{M_{\saturn}R^5}{a^3}\left(k_f-k_2(\nu_2)\right)\sin2\mathcal{M}+\frac{51}{4}n^2e^2\frac{M_{\saturn}R^5}{a^3}\left(\frac{k_2}{Q}\right)(\nu_2)\cos2\mathcal{M}, \label{eq:Gammadevdis}
\end{eqnarray}
that should be equated with:

\begin{eqnarray}
\Gamma & = & \left(\frac{2}{5}MR^2+\frac{M_{\saturn}R^5}{a^3} \left(k_f\left(\frac{5}{9}+\frac{1}{2}e^2\right)+ek_2(\nu_1)\cos\mathcal{M}+\frac{3}{2}e^2k_2(\nu_2)\cos2\mathcal{M}\right)\right. \nonumber \\
& & +\left.e\left(\frac{k_2}{Q}\right)(\nu_1)\sin\mathcal{M}+\frac{3}{2}e^2\left(\frac{k_2}{Q}\right)(\nu_2)\sin2\mathcal{M}\right)\ddot{\sigma} \nonumber \\
& & -\frac{M_{\saturn}R^5}{a^3}(n-\dot{\varpi})\left(k_2(\nu_1)e\sin\mathcal{M}+3k_2(\nu_2)e^2\sin2\mathcal{M}\right. \nonumber \\
& & \left.-\left(\frac{k_2}{Q}\right)(\nu_1)e\cos\mathcal{M}-3\left(\frac{k_2}{Q}\right)(\nu_2)e^2\cos2\mathcal{M}\right)\dot{\sigma},\label{eq:thcinetiq2dis}
\end{eqnarray}
to give the following equation, ruling the libration of the satellite:

\begin{eqnarray}
& & K_0+K_1\sin2\sigma+K_2\sin(\mathcal{M}+2\sigma)+K_3\sin(\mathcal{M}-2\sigma)+K_4\sin(2\mathcal{M}-2\sigma) \nonumber \\ 
& & +K_5\sin\mathcal{M}+K_6\sin2\mathcal{M}+K_{12}\cos\mathcal{M}+K_{13}\cos2\mathcal{M}  \nonumber \\
& = & K_7\ddot{\sigma}+K_8\ddot{\sigma}\cos\mathcal{M}+K_9\ddot{\sigma}\cos2\mathcal{M}+K_{10}\dot{\sigma}\sin\mathcal{M}+K_{11}\dot{\sigma}\sin2\mathcal{M} \nonumber \\
& & +K_{14}\ddot{\sigma}\sin\mathcal{M}+K_{15}\ddot{\sigma}\sin2\mathcal{M}+K_{16}\dot{\sigma}\cos\mathcal{M}+K_{17}\dot{\sigma}\cos2\mathcal{M}, \label{eq:biglibradis}
\end{eqnarray}
with

\begin{eqnarray}
K_0    & = & 18n^2e^2\frac{M_{\saturn}R^5}{a^3}\left(\frac{k_2}{Q}\right)(\nu_1), \label{eq:K0} \\
K_{12} & = & 6en^2\frac{M_{\saturn}R^5}{a^3}\left(\frac{k_2}{Q}\right)(\nu_1), \label{eq:K12} \\
K_{13} & = & \frac{51}{4}e^2n^2\frac{M_{\saturn}R^5}{a^3}\left(\frac{k_2}{Q}\right)(\nu_2), \label{eq:K13} \\
K_{14} & = & \frac{M_{\saturn}R^5}{a^3}e\left(\frac{k_2}{Q}\right)(\nu_1), \label{eq:K14} \\
K_{15} & = & \frac{3}{2}\frac{M_{\saturn}R^5}{a^3}e^2\left(\frac{k_2}{Q}\right)(\nu_2), \label{eq:K15} \\
K_{16} & = & \frac{M_{\saturn}R^5}{a^3}(n-\dot{\varpi})e\left(\frac{k_2}{Q}\right)(\nu_1), \label{eq:K16} \\
K_{17} & = & 3\frac{M_{\saturn}R^5}{a^3}(n-\dot{\varpi})e^2\left(\frac{k_2}{Q}\right)(\nu_2), \label{eq:K17}
\end{eqnarray}
the other $K_i$ being the same as in the conservative case (Eq.~\ref{eq:K1} to \ref{eq:K11}).

\subsection{Librational solution}

\par As before, the Eq.(\ref{eq:biglibradis}) can be simplified into

\begin{equation}
\label{eq:smalllibradis}
\ddot{\sigma}+\omega_0^2\sigma=\kappa_0+\kappa_1\sin\mathcal{M}+\kappa_2\cos\mathcal{M},
\end{equation}
with

\begin{eqnarray}
\kappa_0 & = & \frac{18e^2M_{\saturn}\frac{R^5}{a^3}\left(\frac{k_2}{Q}\right)(\nu_1)}{\frac{2}{5}MR^2+\frac{M_{\saturn}R^5}{a^3}k_f\left(\frac{5}{9}+\frac{e^2}{2}\right)}, \label{eq:J0} \\
\kappa_2 & = & \frac{6eM_{\saturn}\frac{R^5}{a^3}\left(\frac{k_2}{Q}\right)(\nu_1)}{\frac{2}{5}MR^2+\frac{M_{\saturn}R^5}{a^3}k_f\left(\frac{5}{9}+\frac{e^2}{2}\right)}, \label{eq:J2} \\
\end{eqnarray}
$\omega_0$ and $\kappa_1$ being defined as before (Eq.\ref{eq:omega02} \& \ref{eq:J1}). This results in:

\begin{equation}
\label{eq:smallsigmadis}
\sigma(t) = \frac{\kappa_0}{\omega_0^2}+\frac{\kappa_1}{\omega_0^2-(n-\dot{\varpi})^2}\sin\mathcal{M}+\frac{\kappa_2}{\omega_0^2-(n-\dot{\varpi})^2}\cos\mathcal{M}.
\end{equation}
This formula can be also written as

\begin{equation}
\label{eq:smallsigmadis2}
\sigma(t) = \sigma_0+\sigma_1\sin(\mathcal{M}+\phi_1),
\end{equation}
with 

\begin{eqnarray}
\sigma_0 & = & \frac{6e^2M_{\saturn}\frac{R^5}{a^3}\left(\frac{k_2}{Q}\right)(\nu_1)}{\left(1-\frac{5}{2}e^2\right)(I_{22}-I_{11})^{(f)}}, \label{eq:S0} \\
\sigma_1 & = & \frac{\sqrt{\kappa_1^2+\kappa_3^2}}{\omega_0^2-(n-\dot{\varpi})^2}, \label{eq:S1} \\
\tan\phi_1 & = & \frac{M_{\saturn}\frac{R^5}{a^3}\left(\frac{k_2}{Q}\right)(\nu_1)}{(I_{22}-I_{11})^{(f)}+M_{\saturn}\frac{R^5}{a^3}\left(k_f-k_2(\nu_1)\right)}. \label{eq:tanphi2}
\end{eqnarray}

\par We can see that the dissipative tide shifts the phase of the libration, not only in the periodic term, but also in adding a constant term that corresponds to the expected lag.

\par Until recently, this dissipation was considered only for the Moon. \citet{wbyrd2001} denoted the libration as $\tau$ and the constant lag $\sigma_0$ as $\Delta\tau$, 
its formulation being given by the Eq.~(34a) in \emph{Ibid.}. Their formulation is equivalent to mine provided that

\begin{enumerate}

\item the expansion is limited to the order 2 in the eccentricity,

\item their $k_2/Q$ is my $(k_2/Q)(\nu_1)$,

\item their mean triaxiality is only due a frozen component, i.e. their $(B-A)$ is my $(I_{22}-I_{11})^{(f)}$.

\end{enumerate}
The presence of cosines in the librations because of the dissipative tide has also been pointed out in \citep{mfd2016}. This study presented a formula
very similar to Eq.~(\ref{eq:smallsigmadis}), the amplitudes being denoted as $\alpha_s$ and $\beta_s$.

\subsection{Numerical validation}

\par
\begin{figure}[ht]
\centering
\begin{tabular}{cc}
\includegraphics[width=8cm,height=5.5cm]{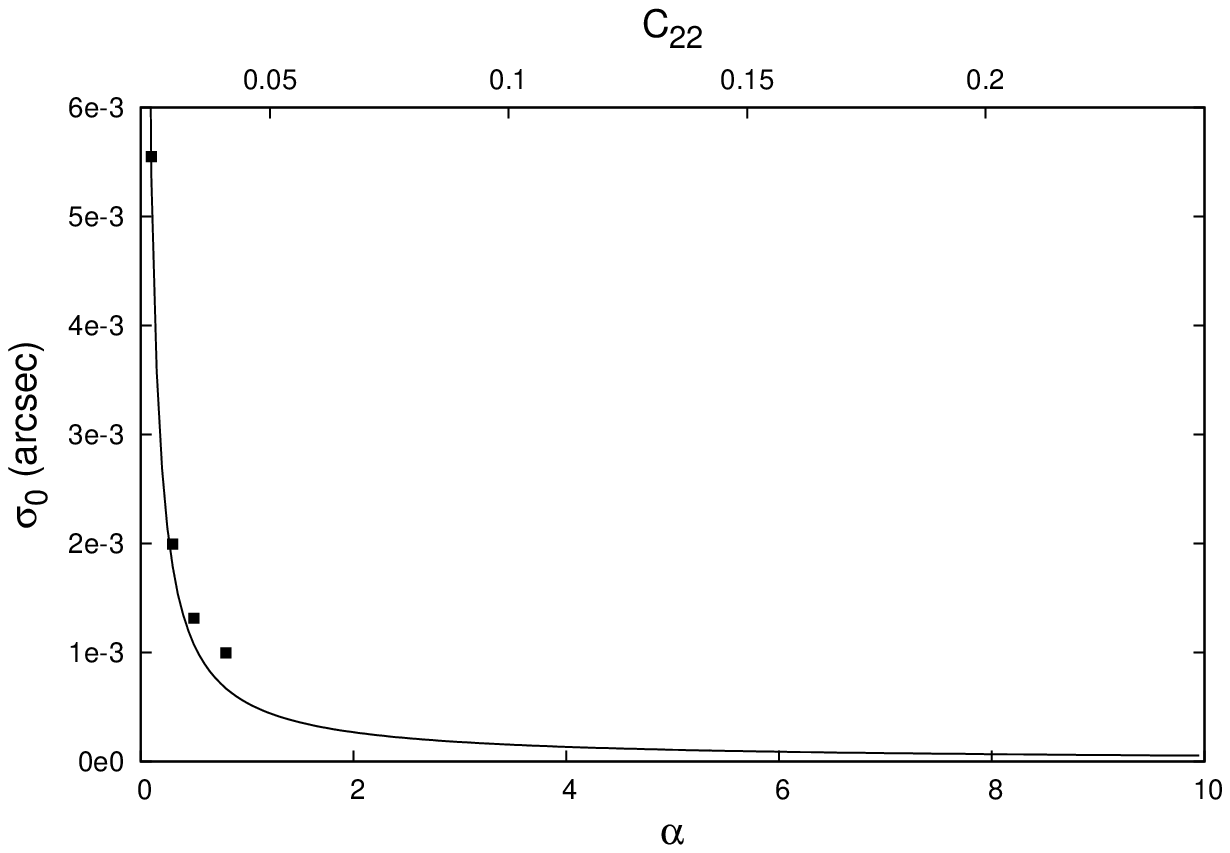} & \includegraphics[width=8cm,height=5.5cm]{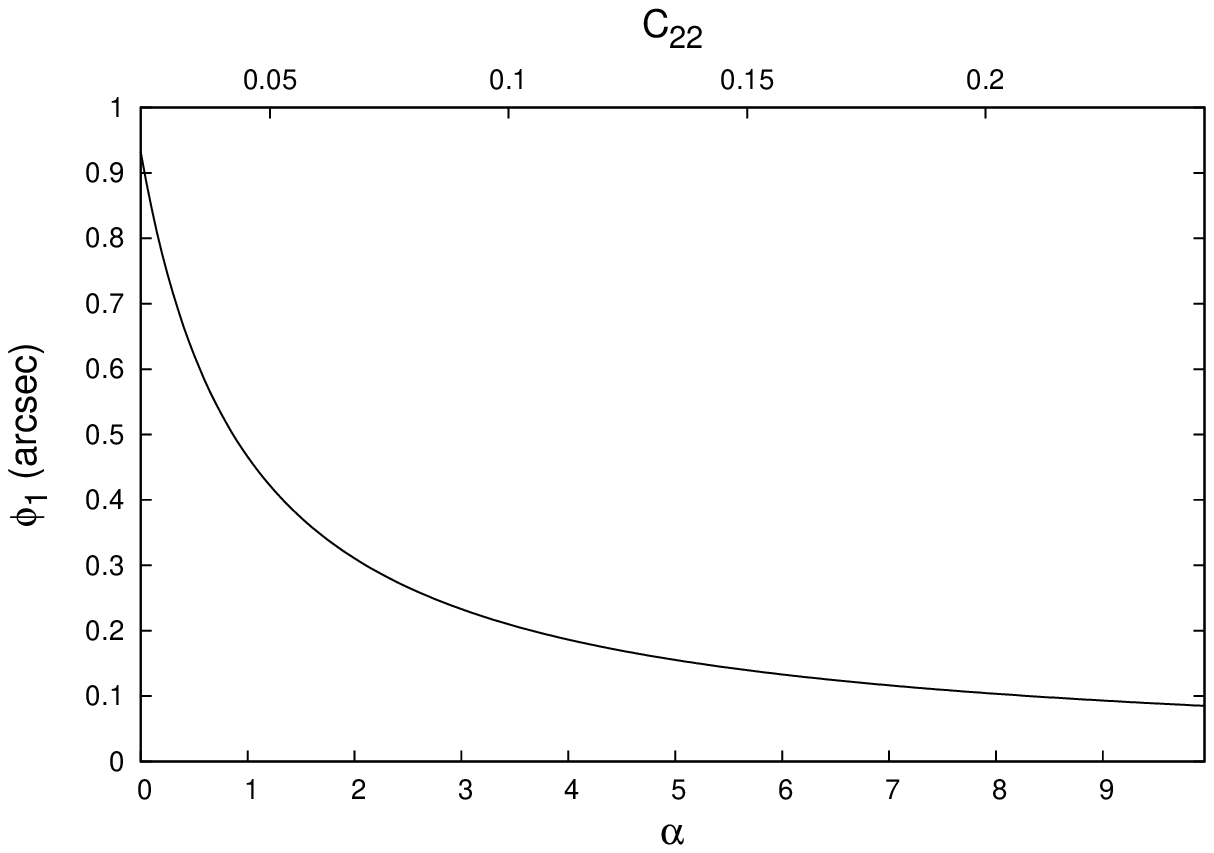} \\
\includegraphics[width=8cm,height=5.5cm]{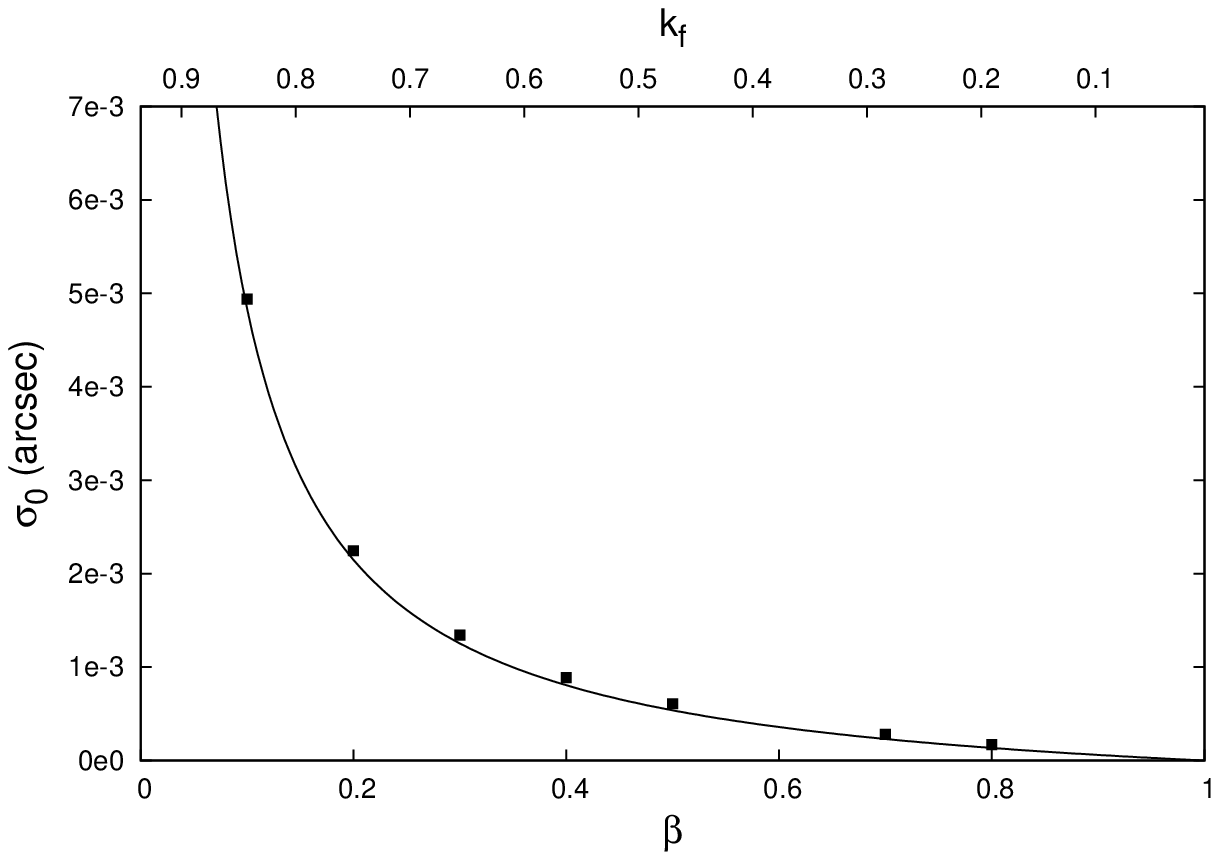} & \includegraphics[width=8cm,height=5.5cm]{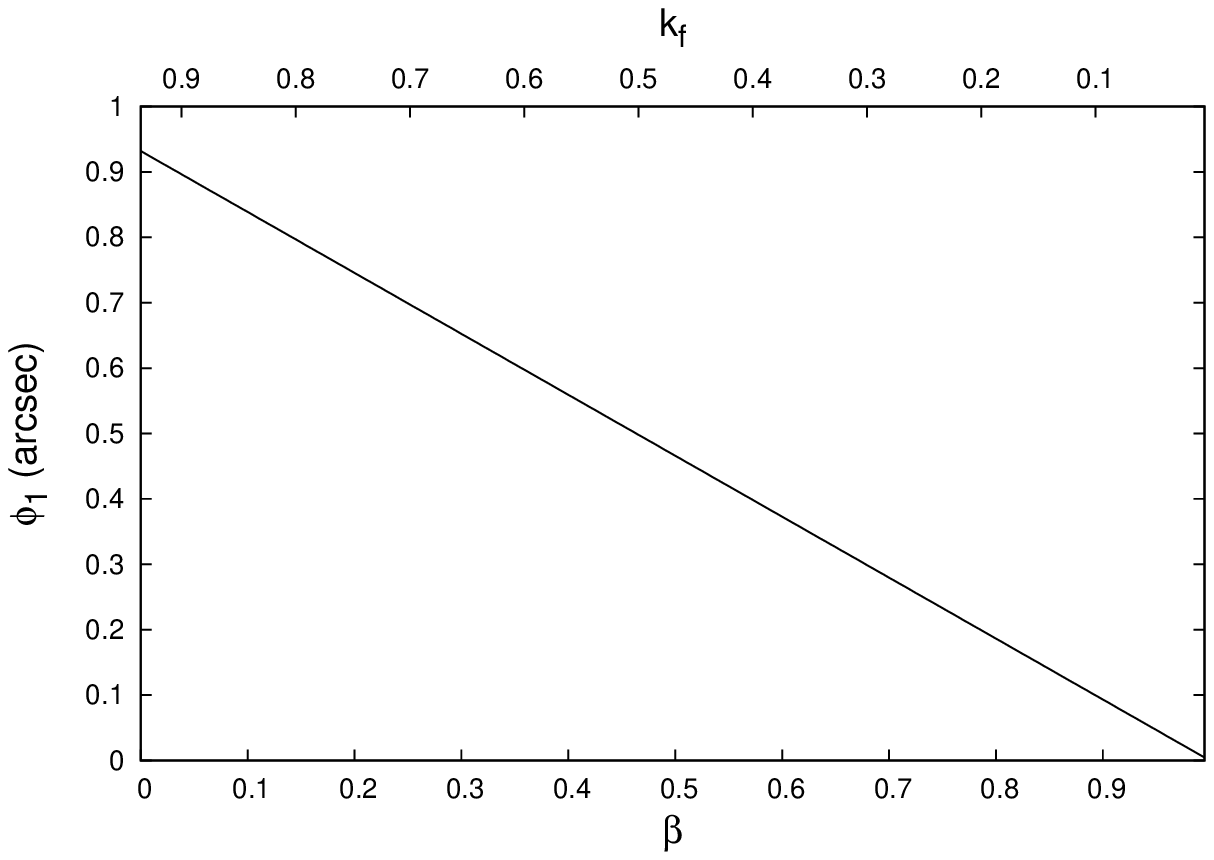}
\end{tabular}
\caption[Rotational quantities for Epimetheus in the dissipative case]{Rotational quantities for Epimetheus in the dissipative case, for $k_f=1.5$
(top), and $C_{22}=1.426\times10^{-2}$ (down). The lines come from the analytical formulae, while the squares result from numerical simulations.\label{fig:epimdis}}
\end{figure}

\begin{figure}[ht]
\centering
\begin{tabular}{cc}
\includegraphics[width=8cm,height=5.5cm]{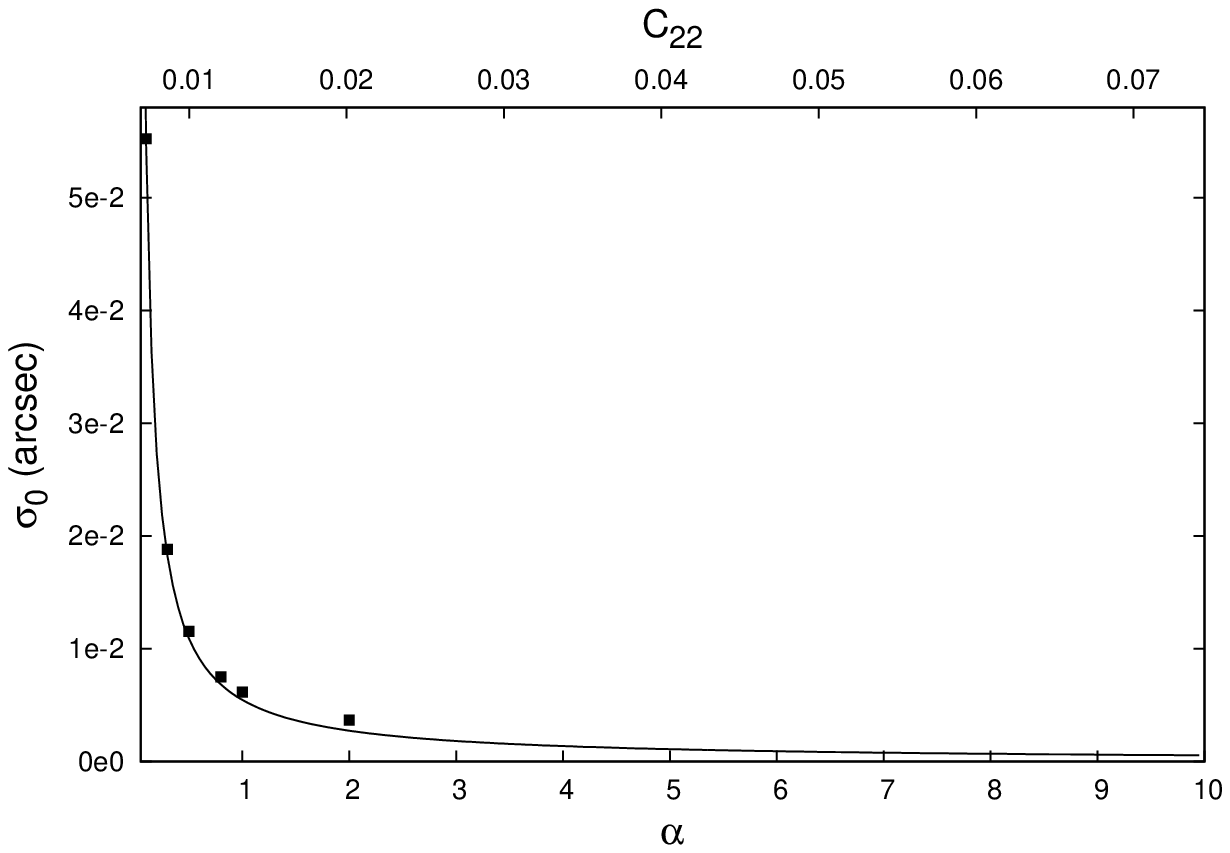} & \includegraphics[width=8cm,height=5.5cm]{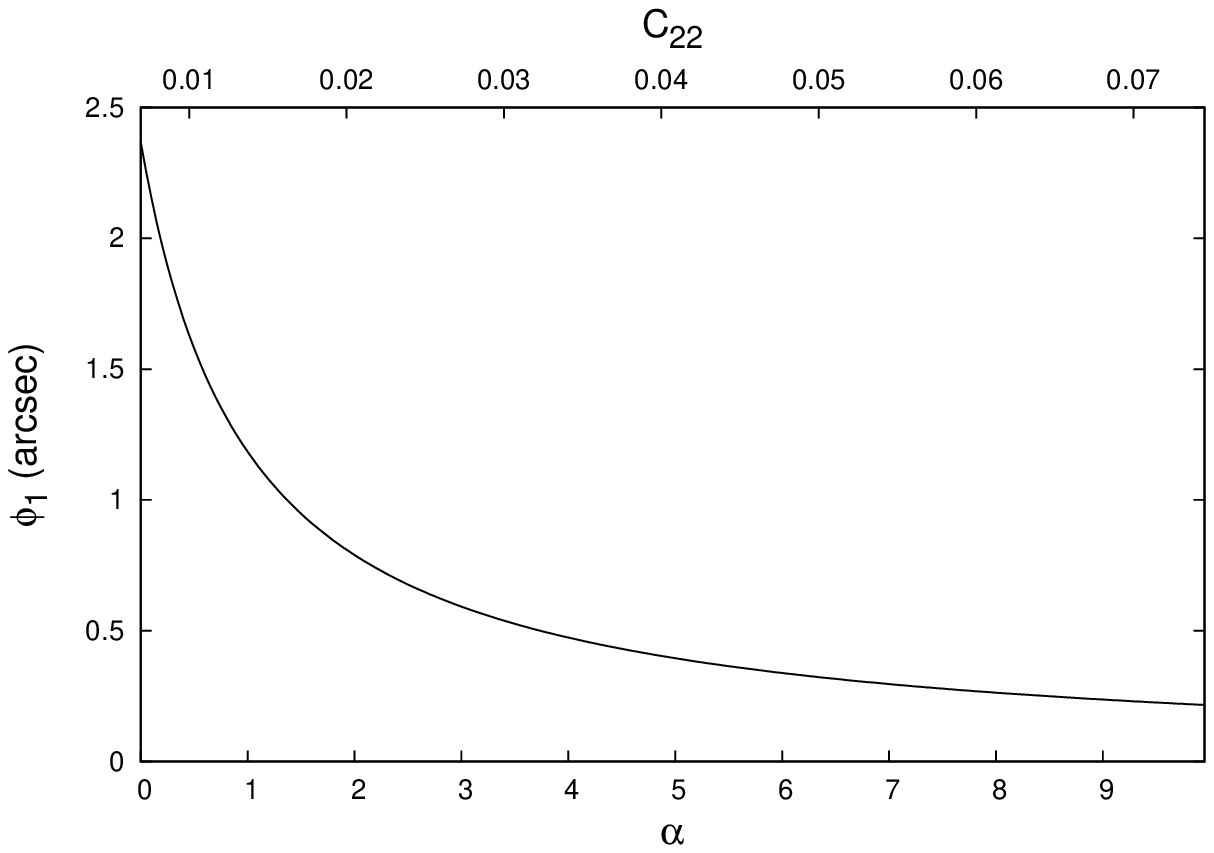} \\
\includegraphics[width=8cm,height=5.5cm]{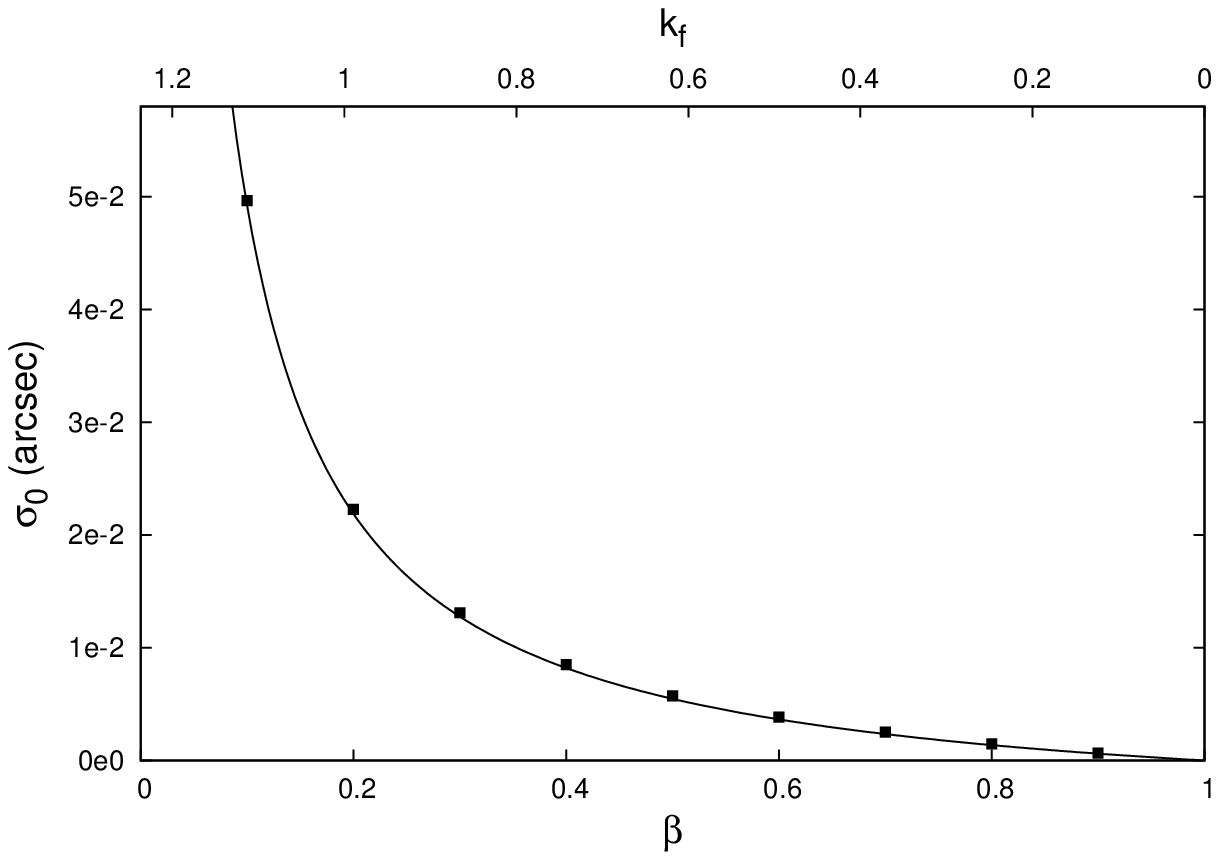} & \includegraphics[width=8cm,height=5.5cm]{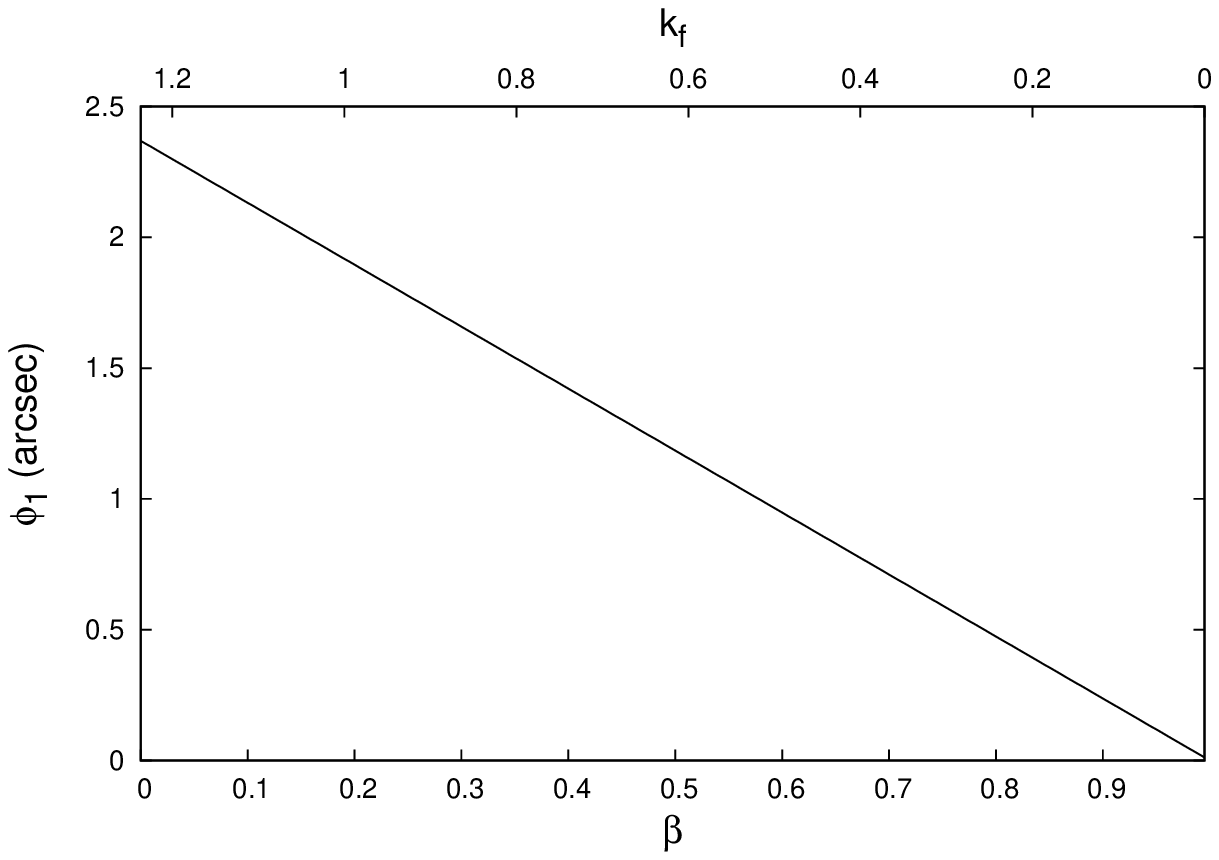}
\end{tabular}
\caption[Rotational quantities for Mimas in the dissipative case]{Rotational quantities for Mimas in the dissipative case, for $k_f=1.5$ 
(top), and $C_{22}=5.606\times10^{-3}$ (down). The lines come from the analytical formulae, while the squares result from numerical simulations.\label{fig:mimasdis}}
\end{figure}

 Similarly to before, I present numerical simulations of the librations of Epimetheus (Fig.~\ref{fig:epimdis}) and Mimas (Fig.~\ref{fig:mimasdis}), 
from the Eq.~(\ref{eq:biglibradis}). Here, only phases shifts are depicted, but the validity of the numerical integrations has been  also successfully tested 
on the frequency of the proper oscillations $\omega_0$ and the diurnal amplitude $\sigma_1$, which does not significantly differ from the \emph{Amplitude 1} in 
the Fig.~\ref{fig:epimkf} to \ref{fig:mimasC22}.

\par The permanent phase lag $\sigma_0$ is numerically determined from the constant term in the frequency analysis of the variable $\sigma(t)$ when identified, while 
the shift in the diurnal libration $\phi_1$ should be extracted from the phase of the term oscillating at the frequency $n-\dot{\varpi}$. Unfortunately, I do not
have enough accuracy on this phase to get a numerical determination of $\phi_1$ with enough confidence, this is why only the analytical result is shown.
The plots suggest a good confidence for the expected value of $\sigma_0$. I might actually have an error of a few percents.

\placefigure{fig:epimdis}

\placefigure{fig:mimasdis}

\par No shift has been determined for Epimetheus. The measurements of the librations of Mimas by \citet[Tab.~1]{trlcrrn2014} suggest $\phi_1=(6.35\pm0.8)^{\circ}$, 
which is too large to be explained with the Maxwell rheology and a Maxwell time of $5$ days. This is why I now investigate the influence of the rheology.

\section{Influence of the rheology\label{sec:rheology}}

\par
\begin{figure}[ht]
\centering
\begin{tabular}{cc}
\includegraphics[width=8cm,height=5.5cm]{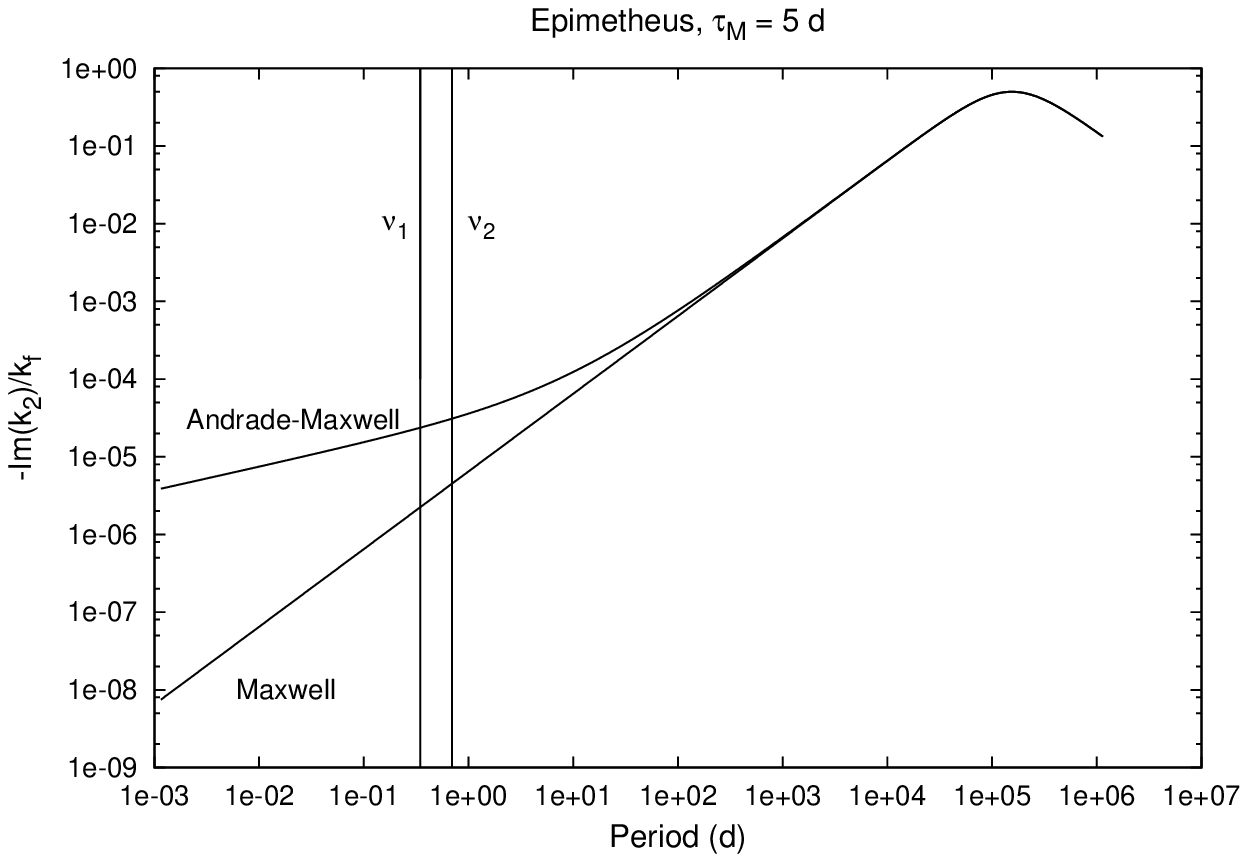} & \includegraphics[width=8cm,height=5.5cm]{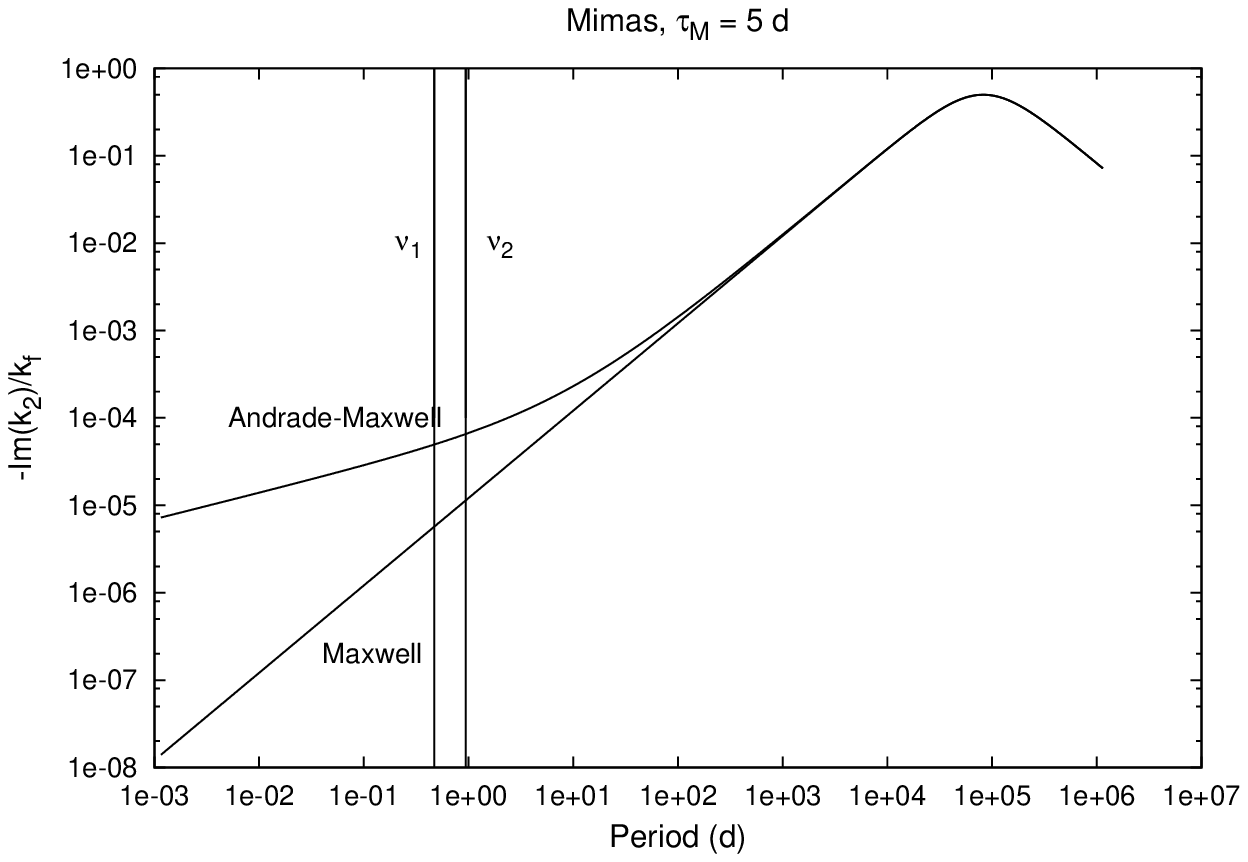}
\end{tabular}
\caption[Dissipative tides]{Dissipative part of the tides, for the Maxwell and Andrade-Maxwell models. We can see a significant discrepancy for high-frequency excitations, 
which affects the quantity $k_2/Q$ at the diurnal and semi-diurnal frequencies.\label{fig:imk2}}
\end{figure}

 The Maxwell-Andrade model gives:

\begin{equation}
\label{eq:Imk2Efroimsky}
-\mathfrak{I}(k_2^*)  =  k_f\frac{A_2\mathcal{B}}{\mathcal{A}^2+2\mathcal{A}A_2+A_2^2+\mathcal{B}^2}, 
\end{equation}
the quantitites $A_2$, $\mathcal{A}$ and $\mathcal{B}$ having already been defined (Eq.\ref{eq:A2}, \ref{eq:Aronde} \& \ref{eq:Bronde}). As shown in the
Fig.~\ref{fig:imk2}, this model gives a higher dissipation than the Maxwell model at high frequencies.

\placefigure{fig:imk2}

\par
\begin{figure}[ht]
\centering
\begin{tabular}{cc}
\includegraphics[width=8cm,height=5.5cm]{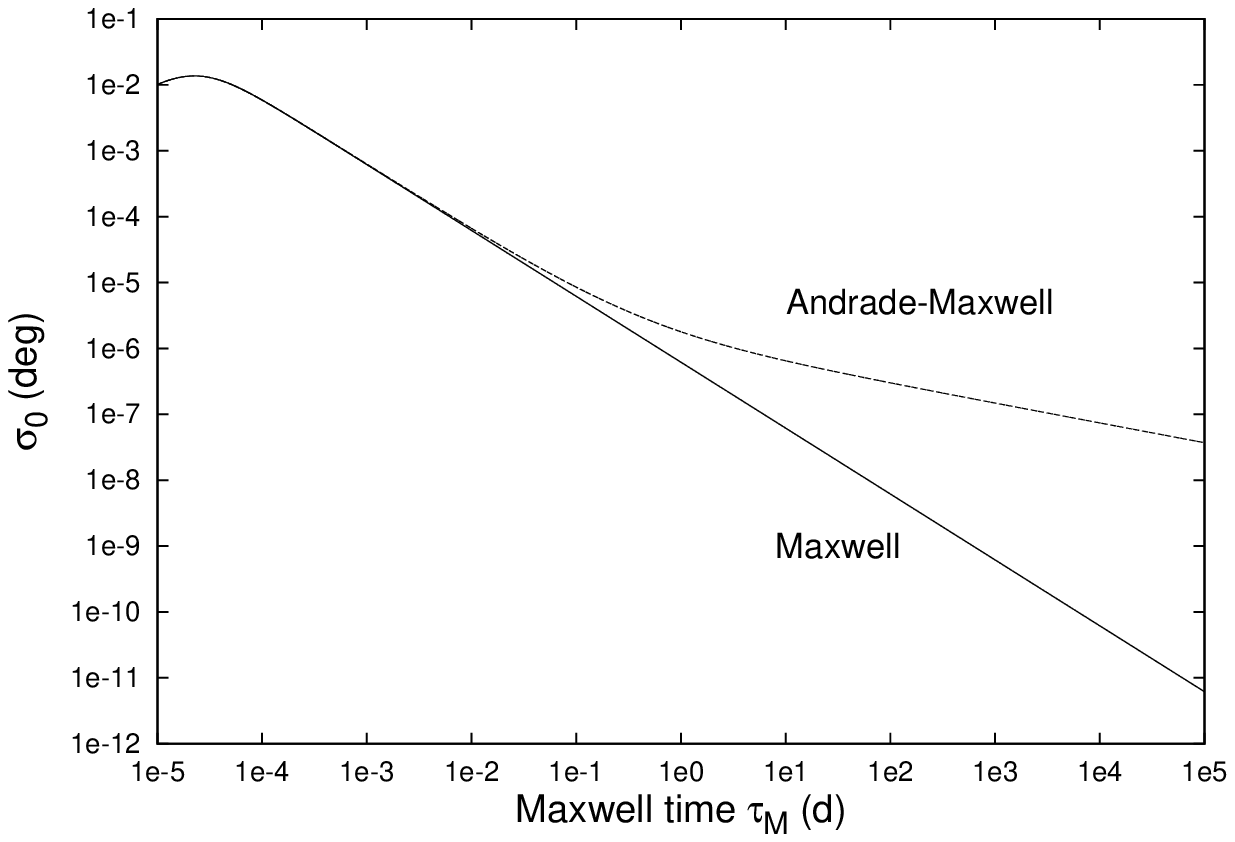} & \includegraphics[width=8cm,height=5.5cm]{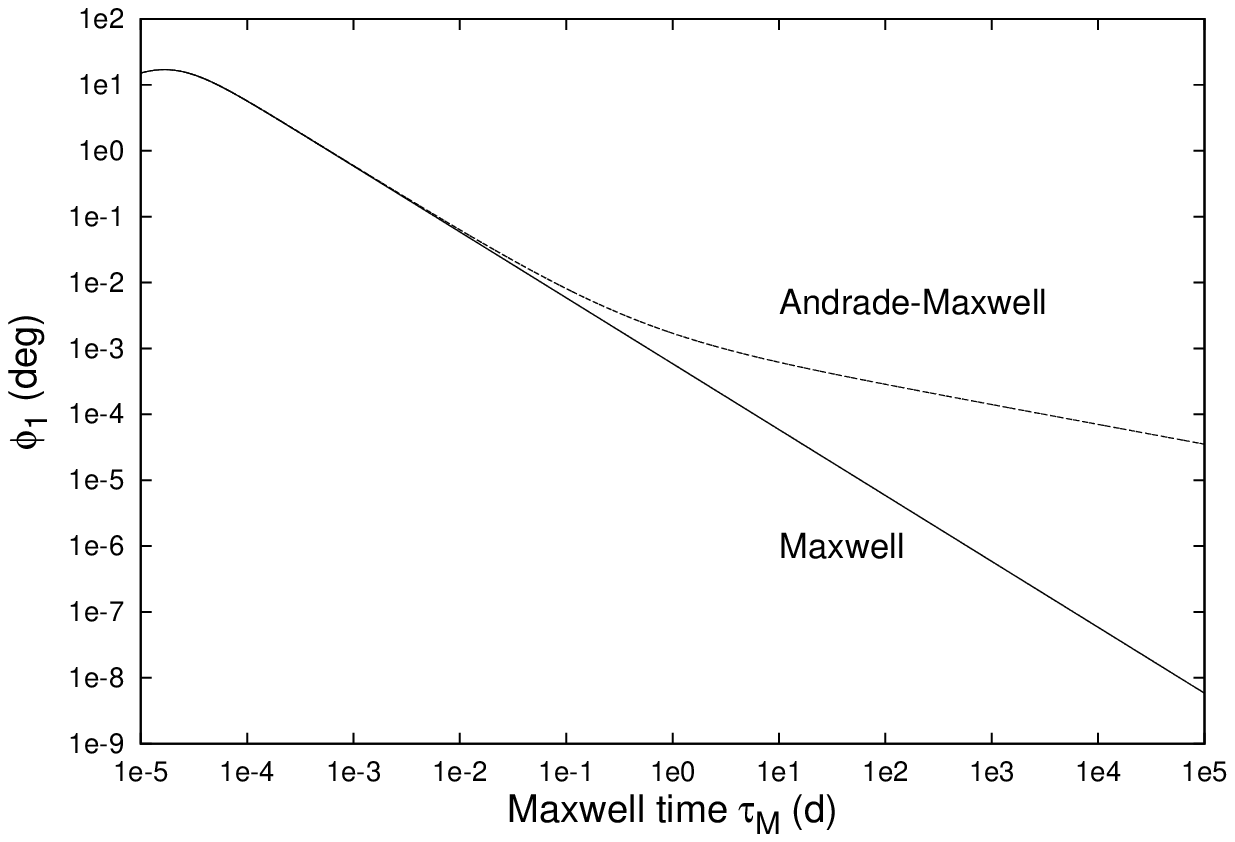}
\end{tabular}
\caption[Influence of the rheology on the rotation of Epimetheus]{Influence of the rheology on the rotation of Epimetheus, for $\alpha = 1.21$. The Andrade-Maxwell model has been applied with $N=0.3$, this parameter influencing the slope for high Maxwell times. \label{fig:rheoepim}}
\end{figure}

\begin{figure}[ht]
\centering
\begin{tabular}{cc}
\includegraphics[width=8cm,height=5.5cm]{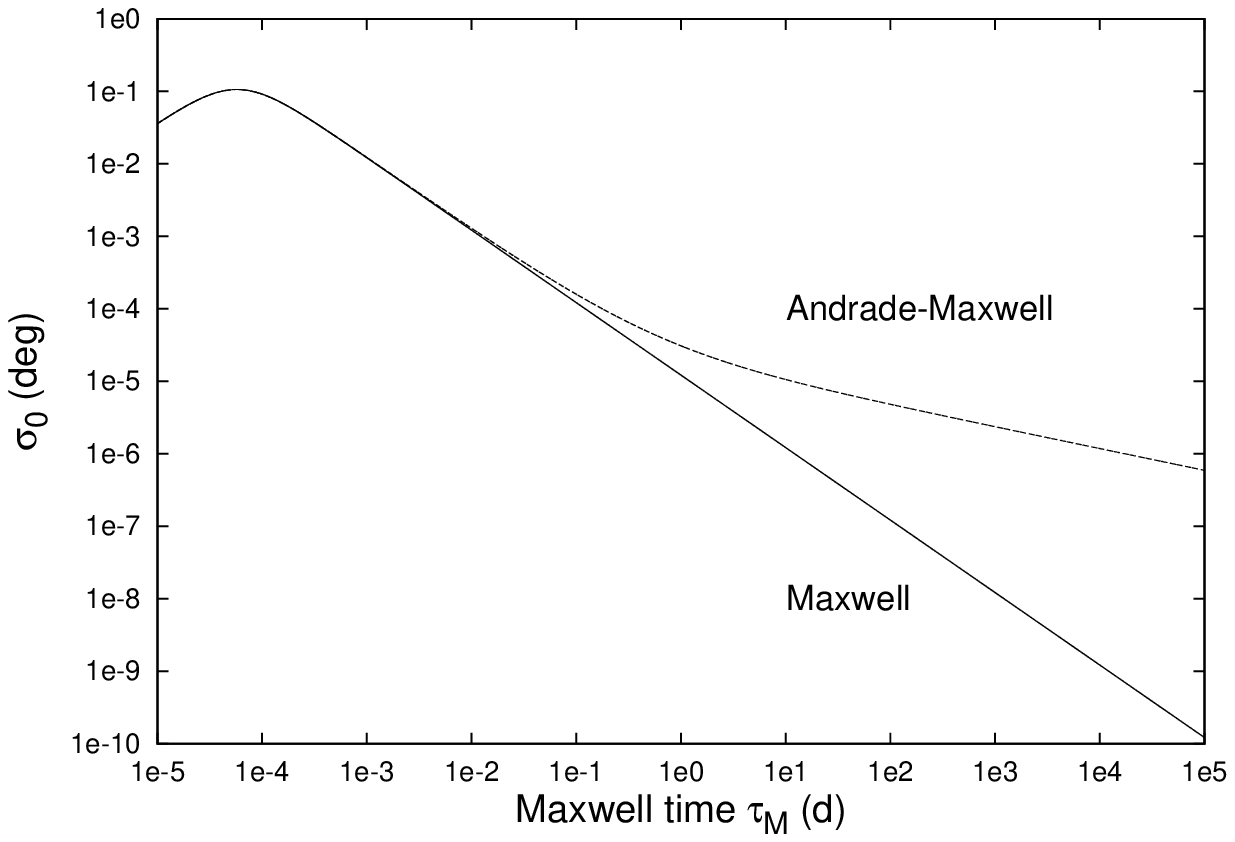} & \includegraphics[width=8cm,height=5.5cm]{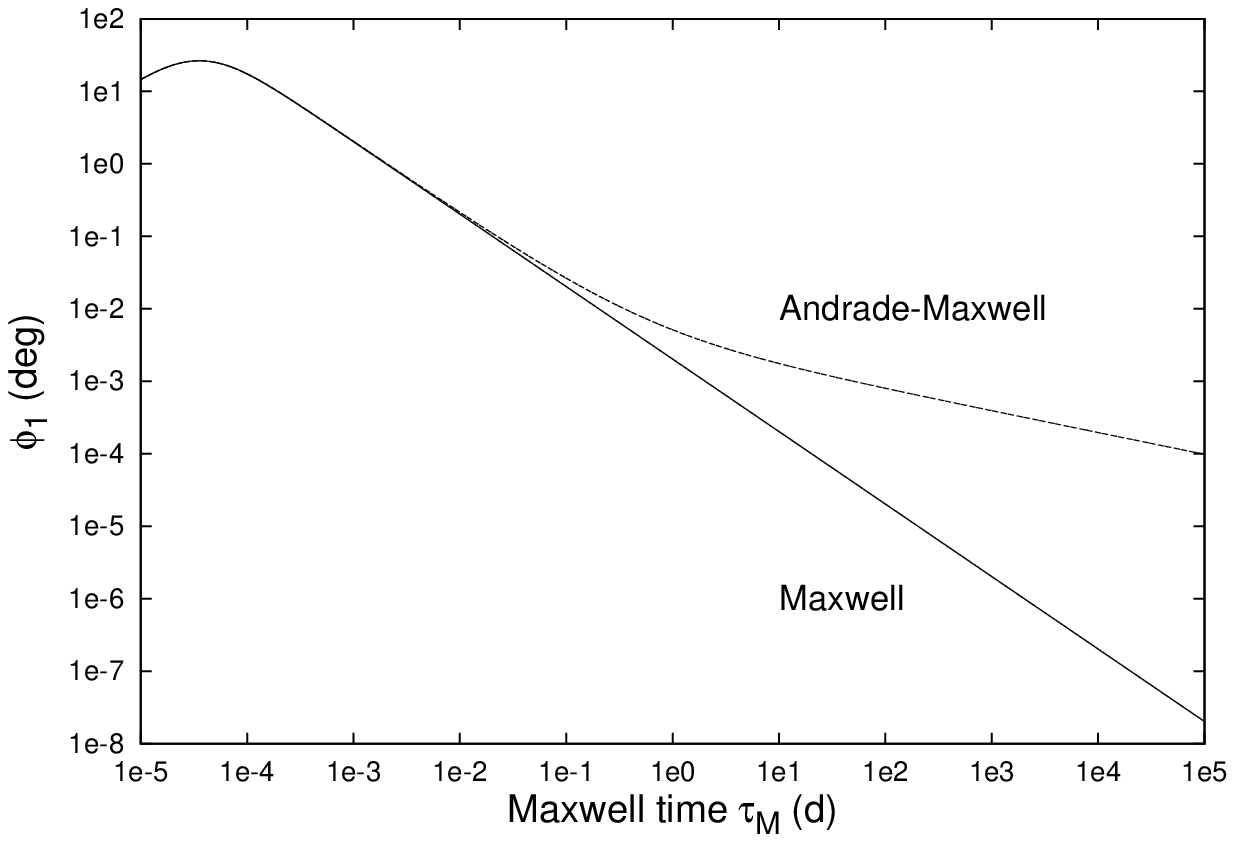}
\end{tabular}
\caption[Influence of the rheology on the rotation of Mimas]{Influence of the rheology on the rotation of Mimas, for $\alpha = 0.6225$, and $N=0.3$. \label{fig:rheomimas}}
\end{figure}

 I here compare the phase shifts given by the Maxwell and Andrade-Maxwell rheologies, for different Maxwell times (Fig.~\ref{fig:rheoepim} \& \ref{fig:rheomimas}).

\placefigure{fig:rheoepim}

\par The phase shifts $\sigma_0$ and $\phi_1$ have been plotted for Epimetheus for $\alpha = 1.21$, i.e. consistently with the measured libration. We see that the 2 rheological models 
diverge only for large Maxwell times. The reason for this is that the Andrade-Maxwell rheology considers a high-frequency effect, i.e. the Andrade creep, while the quantities I depict
depend on $k_2(\nu_1)$. So, a difference between these 2 models should appear only when the Maxwell time is large with respect to the orbital period. We also see that the phase shifts
should be very small, except for Maxwell times that are very short, i.e. a few seconds. This could explain why neither $\sigma_0$  nor $\phi_1$ has been detected for Epimetheus.

\placefigure{fig:rheomimas}

\par
\begin{figure}[ht]
\centering
\begin{tabular}{cc}
\includegraphics[width=8cm,height=5.5cm]{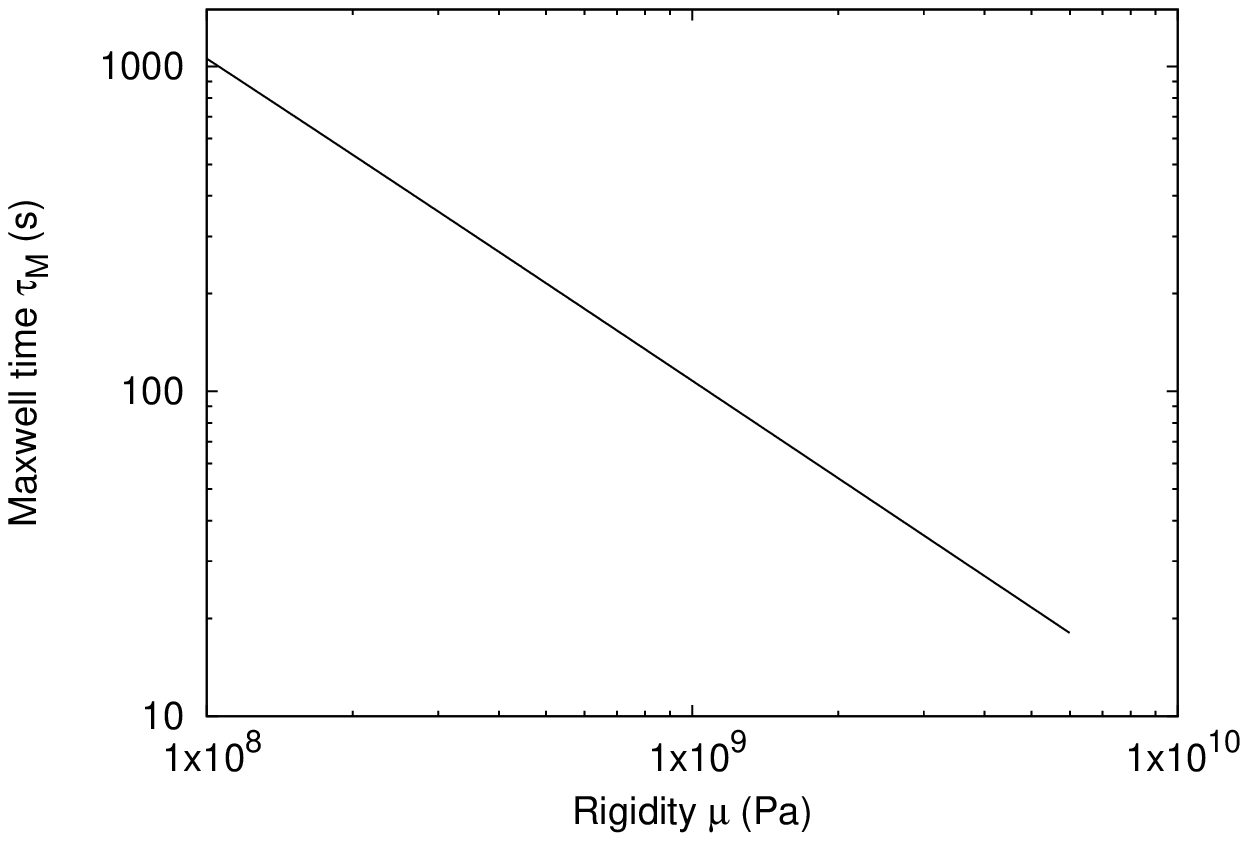} & \includegraphics[width=8cm,height=5.5cm]{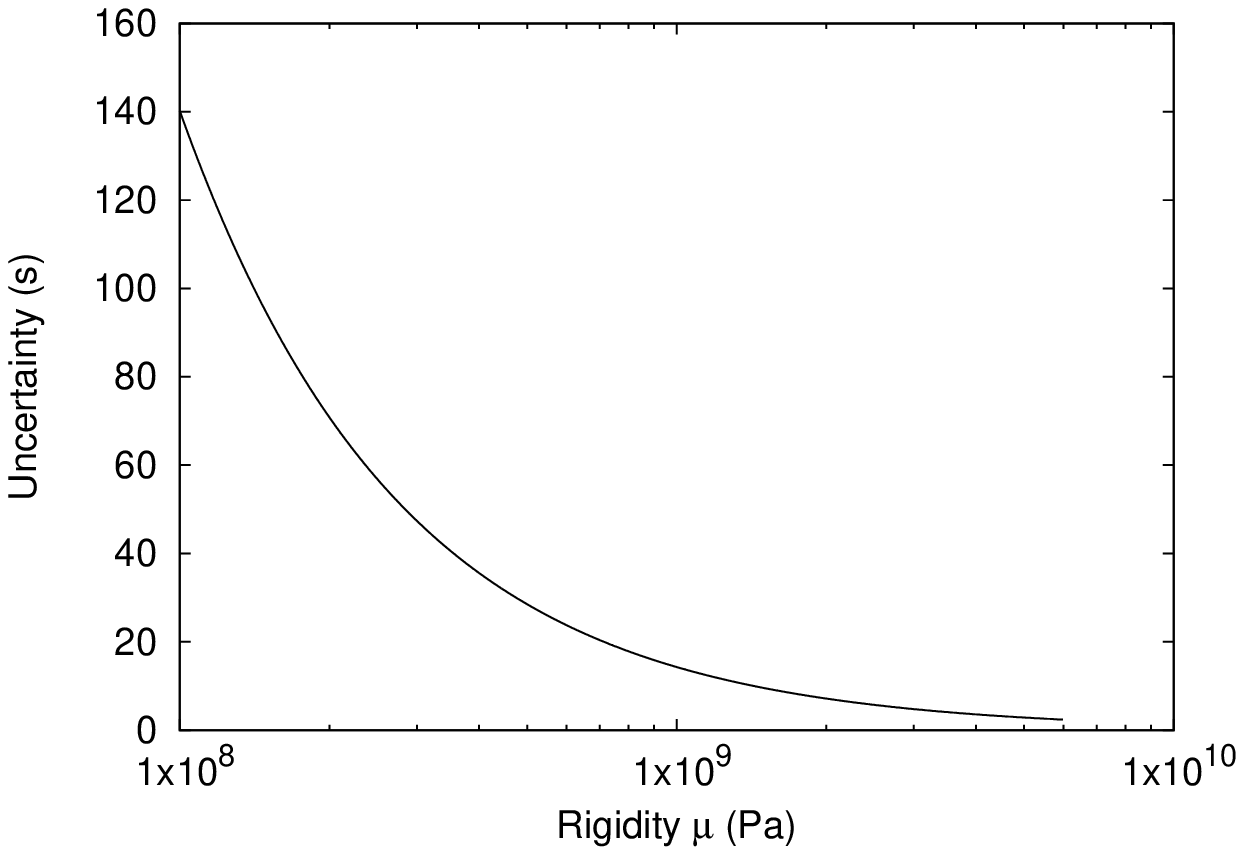}
\end{tabular}
\caption[Deduced Maxwell times for Mimas]{Maxwell times of Mimas deduced from the diurnal phase shift $\phi_1$ (left), and the uncertainty associated (right). These numbers are much shorter than 
expected from our knowledge of Mimas.\label{fig:maxwellmimas}}
\end{figure}

 As for Epimetheus, the Fig.~\ref{fig:rheomimas} suggests that the phase shifts should be very small for Mimas. However, \citet{trlcrrn2014} have measured $\phi_1 = 6.35\pm0.8^{\circ}$,
which suggests $\tau_M = 27.15_{-3.15}^{+4.05}$ s for a rigidity $\mu=4$ GPa. A rigidity ten times smaller, i.e. $\mu=0.4$ GPa, would give $\tau_M=270_{-31}^{+40}$ s 
(Fig.~\ref{fig:maxwellmimas}). A smaller $k_f$ would give a Maxwell time of the same order of magnitude, since it would be partly counterbalanced by a larger $\alpha$. For instance, 
we would have $\alpha=1.295\pm0.033$ and $\tau_M=18.62_{-2.23}^{+2.96}$ s for a rigidity $\mu=4$ GPa and $k_f=1$. These are very small times, if we keep in mind that a Maxwell time of 
the magnitude of the day is to be expected for ice near its melting point \citep[e.g.]{nm2009}, and larger times for denser materials. Actually, no realistic rigidity results in a realistic Maxwell time.

\placefigure{fig:maxwellmimas}

\par This result discards a model of solid structure for Mimas. In \citet{trlcrrn2014}, the authors retained two internal structures which could explain the measured 
amplitude of libration: a solid body with an excess of triaxiality due to a dense and elongated core, and a global ocean. The phase shift seems to speak for the global ocean.

\section{Conclusion\label{sec:conclusion}}

\par In this study I have proposed a novel theory of librations of a synchronous satellite, in which the role of the frozen component of the inertia of a triaxial satellite 
is differentiated from the one of the secular elastic deformation. The frozen component should not be addressed as a result of the hydrostatic equilibrium, a recent study
by \citet{vbt2016} mentions this issue. I have shown that
the measured physical librations of Epimetheus and Mimas are consistent with the presence of these two components in the inertia of the bodies. However,
the introduction of the dissipative tides introduces a phase shift in the librations, which is far too small to explain the one which has been measured for
Mimas. This strenghtens the assessment of a global ocean.

\par The detection of the phase shift in the librations of Mimas is a key result, which should encourage its investigation for other bodies, because it 
probably contains information on the ocean. For that, this study should be extended to a more complex structure, in which a solid crust is partly decoupled
from a rigid core by a global fluid layer.

\par Features at the surface of Europa have been interpreted as an evidence for supersynchronous rotation \citep{gghhmptosgbdcbv1998}, the present study cannot 
explain it, but an influence of the interior is not to be discarded.

\section*{Acknowledgments}

\par This study took benefit from the financial support of the contract Prodex CR90253 from the Belgian Science Policy Office (BELSPO), and is part of the 
activities of the ISSI Team \emph{Constraining the dynamical timescale and internal processes of the Saturn System from astrometry}. I am indebted to
Rose-Marie Baland, Marie Yseboodt, and two anonymous reviewers, for their careful readings and valuable suggestions.

\appendix

\section{Notations used in this paper}

\placetable{tab:notations}

\clearpage

\listoftables

\clearpage

\begin{table}[ht]
\centering
\caption[Orbital dynamics and gravity field of the main satellites of Saturn]{Orbital dynamics and gravity field of the main satellites of Saturn. The orbital data
are mean Keplerian elements given by the JPL HORIZON portal, except for the eccentricity of Enceladus that is taken from \citep{vd1995}, and the mean motions
of Janus and Epimetheus, taken from \citep{n2010}. The masses come from \citep{jabcijmmpporrs2006,jspbcem2008}, and the Stokes coefficients from 
\citep{isphjlnaadt2014,tzhnjip2016,irjrstaa2010}. Two gravity solutions have been published for Rhea and Titan.\label{tab:gravity}}
\begin{tabular}{l|ccccc}
Satellites & $\mathcal{G}M$ & $n$ & $e$ & $J_2$ & $C_{22}$ \\
 & (km$^3$.s$^{-2}$) & (rad/y) & $(\times 100)$ & $(\times 10^5)$ & $(\times 10^5)$ \\
\hline
Janus & $0.12660$ & $3304.014328$ & $0.68$ & -- & -- \\ 
Epimetheus & $0.03513$ & $3304.014328$ & $0.98$ & -- & -- \\
Mimas & $2.5023$ & $2435.144271$ & $1.96$ & -- & -- \\
Enceladus & $7.2096$ & $1674.867267$ & $0.5$ & $543.25$ & $154.98$ \\
Tethys & $41.2097$ & $1215.663921$ & $0.01$ & -- & -- \\
Dione & $73.1127$ & $838.510861$ & $0.22$ & -- & -- \\
Rhea (SOL 1) & $153.9416$ & $508.009307$ & $0.02$ & $94.60$ & $24.21$ \\
(SOL 2) & & & & $95.70$ & $22.70$ \\
Titan (SOL 1) & $8978.1356$ & $143.924045$ & $2.88$ & $3.1808$ & $0.9983$ \\
(SOL 2) & & & & $3.3462$ & $1.0022$ \\
\end{tabular}
\end{table}

\begin{table}[ht]
\centering
\caption{Shapes of the main satellites of Saturn\label{tab:shapes}}
\begin{tabular}{l|cc|l|c}
Satellites & $\mathfrak{a}\times \mathfrak{b} \times \mathfrak{c}$ & $R$ & References & $q$ \\
 & (km$\times$ km$\times$ km) & (km) & & $(\times 10^3)$ \\
\hline
Janus & $101.5\times 92.5\times 76.3$ & $89.5$ & \citet{ttb2009} & $62.07$ \\ 
Epimetheus & $64.9\times 57.0\times 53.1$ & $58.1$ & \citet{ttb2009} & $61.20$ \\
Mimas & $206.60\times 195.73\times 190.47$ & $197.49$ & \citet{trlcrrn2014} & $18.33$ \\
Enceladus & $256.2\times 251.4\times 248.6$ & $252.24$ & \citet{tttbjlhp2016} & $6.270$ \\
Tethys & $538.4\times 528.3\times 526.3$ & $531.0$ & \citet{t2010} & $5.391$ \\
Dione & $563.4\times 561.3\times 559.6$ & $561.4$ & \citet{t2010} & $1.709$ \\
Rhea & $765.0\times 763.1\times 762.4$ & $736.5$ & \citet{t2010} & $0.749$ \\
Titan & $2575.15\times 2574.78\times 2574.47$ & $2574.73$ & \citet{zshlkl2009} & $0.040$ \\
\end{tabular}
\end{table}

\begin{table}[ht]
\centering
\caption[Departures from the hydrostatic equilibrium]{Departures from the hydrostatic equilibrium. The theoretical numbers come from
the Eq.(\ref{eq:hydroshape}) and (\ref{eq:hydrograv}).\label{tab:nonhydro}}
\begin{tabular}{l|ccc|ccc|}
Satellites & \multicolumn{3}{|c|}{$(\mathfrak{b}-\mathfrak{c})/(\mathfrak{a}-\mathfrak{c})$} & \multicolumn{3}{|c|}{$J_2/C_{22}$} \\
 & Measured & Theoretical & Error & Measured & Theoretical & Error \\
\hline
Janus & $0.643$ & $0.113$ & $469.4\%$ & -- & $2.151$ & -- \\
Epimetheus & $0.331$ & $0.115$ & $187.6\%$ & -- & $2.169$ & -- \\
Mimas & $0.326$ & $0.210$ & $55.2\%$ & -- & $2.989$ & -- \\
Enceladus & $0.368$ & $0.236$ & $56.0\%$ & $3.505$ & $3.214$ & $9.1\%$ \\
Tethys & $0.165$ & $0.238$ & $30.6\%$ & -- & $3.231$ & -- \\
Dione & $0.447$ & $0.246$ & $81.7\%$ & -- & $3.301$ & -- \\
Rhea (SOL 1) & $0.269$ & $0.248$ & $8.4\%$ & $3.907$ & $3.319$ & $17.7\%$ \\
(SOL 2) & & & & $4.216$ & & $27.0\%$ \\
Titan (SOL 1) & $0.456$ & $0.251$ & $81.5\%$ & $3.186$ & $3.342$ & $4.7\%$ \\
(SOL 2) &  & & & $3.339$ & & $0.09\%$ \\
\end{tabular}
\end{table}

\begin{table}[ht]
\centering
\caption[Comparison with previous studies]{Comparison with previous studies. The formulae labeled \emph{This study} are given by the Eq.~\ref{eq:omega02} and \ref{eq:J1}, but there expansion
is here limited to the degree 1 in the eccentricity.\label{tab:comparison}}
\begin{tabular}{|c|c|c|c|}
\hline
 & \citet{vbt2013} & \citet{rrc2014} & This study \\
\hline
$\omega_0^2$ & $3n^2\frac{B-A}{C}\left(1-\frac{k_2}{k_f}\right)$ & $3n^2\frac{B-A}{C}$  & $3n^2\frac{(I_{22}-I_{11})^{(f)}}{\frac{2}{5}MR^2+\frac{5}{9}k_fM_{\saturn}\frac{R^5}{a^3}}$ \\
$\kappa_1$ & $6en^2\frac{B-A}{C}\left(1-\frac{5}{6}\frac{k_2}{k_f}\right)$ & $6en^2\left(\frac{B-A}{C}-\frac{h_2}{2}M_{\saturn}\frac{R^5}{Ca^3}\right)$ & $6en^2\frac{(I_{22}-I_{11})^{(f)}+M_{\saturn}\frac{R^5}{a^3}\left(k_f-k_2(\nu_1)\right)}{\frac{2}{5}MR^2+\frac{5}{9}k_fM_{\saturn}\frac{R^5}{a^3}}$ \\
\hline
\end{tabular}
\end{table}

\begin{table}[ht]
\centering
\caption[Physical and dynamical parameters used for Epimetheus and Mimas]{Physical and dynamical parameters used for Epimetheus and Mimas.
The densities $\rho$ and the dynamical parameters $n$, $a$, $e$ and $\dot{\varpi}$ come from JPL HORIZONS, except for the eccentricity of Epimetheus that is the one 
recommended by \citet{ttb2009}. The rigidity $\mu$ of Epimetheus is the one used by \citet{rrc2011}, itself derived from \citep{gs2009} in assuming 
Epimetheus to be a rubble-pile, while the one of Mimas is taken from \citep{mw2008}. The Love numbers are computed from the Maxwell model (Eq.~\ref{eq:rek2}).\label{tab:param}}
\begin{tabular}{|l|l|l|}
\hline
& Epimetheus & Mimas \\
\hline
$\mu$ & $2\times10^{8}$ Pa & $4\times10^{9}$ Pa \\
\hline
$\tau_M$ & $5$ d & $5$ d \\
\hline
$\rho$ & $6.4\times10^2$ kg/m$^3$ & $1.15\times10^3$ kg/m$^3$ \\
\hline
$g$ & $1.04\times10^{-2}$ m/s$^2$ & $6.365\times10^{-2}$ m/s$^2$ \\
\hline
$A_2$ & $4.914\times10^3$ & $2.618\times10^3$ \\
\hline
$k_2(\nu_1)/k_f$ & $2.035\times10^{-4}$ & $3.818\times10^{-4}$ \\
\hline
$n$ & $9.0425$ rad/d & $6.66706$ rad/d \\
\hline
$\dot{\varpi}$ & $3.6\times10^{-2}$ rad/d & $1.745\times10^{-2}$ rad/d \\
\hline
$e$ & $9.8\times10^{-3}$ & $1.96\times10^{-2}$ \\
\hline
$a$ & $151\,453$ km & $185\,565$ km\\
\hline
\end{tabular}
\end{table}

\begin{deluxetable}{lr}
	\tablecolumns{2}
	\tablewidth{0pt}
	\tablecaption{Main notations used in this study.\label{tab:notations}}
	\startdata
 \hline
  $\mathfrak{a}$, $\mathfrak{b}$, $\mathfrak{c}$, $R$				& External radii and mean radius of the satellite \\
  $n$, $a$, $e$									& Mean motion, semimajor axis and eccentricity of the satellite \\
  $\lambda$, $f$, $\mathcal{M}$							& Mean longitude, true longitude and mean anomaly of the satellite \\
  $\varpi$, $\dot{\varpi}$							& Longitude of the pericenter of the satellite, and its precessional rate \\
  $J_2$, $C_{22}$								& Stokes coefficients of the satellite \\
  $M_{\saturn}$, $M$								& Masses of the parent planet (here Saturn) and the satellite \\
  $\rho$, $g$									& Density and surface gravity of the satellite \\
  $X$, $Y$, $Z$									& Cartesian coordinates of a mass element of the satellite \\
  $\mathfrak{r}$, $\theta$, $\ell$						& Spherical coordinates of a mass element of the satellite \\
  $x=x_1$, $y=x_2$, $z=x_3$							& Cartesian coordinates of the unit vector pointing to the planet \\
  $\mathcal{I}^{(f)}$, $\mathcal{I}^{(e)}$, $\mathcal{I}$			& Frozen, elastic, and global tensors of inertia \\
  $\mathcal{I}^{(r)}$, $\mathcal{I}^{(t)}$, $\mathcal{I}^{(v)}$			& Rotational, tidal, and viscoelastic tensors of inertia \\
  $I_{ij}$									& Components of the tensor of inertia \\
  $\tau_M$, $\tau_A$								& Maxwell and Andrade times \\
  $p$, $\sigma = p-\lambda+\pi$							& Rotation angle and argument of the synchronous resonance \\
  $N$										& Andrade parameter \\
  $\mu$, $\eta$ 								& Rigidity and viscosity \\
  \hline
  \enddata
\end{deluxetable}

\clearpage

\listoffigures

\end{document}